\documentclass[twocolumn]{aastex7}  

\usepackage{graphicx}

\graphicspath{{./}}

\usepackage{amsmath}
\usepackage{amssymb}

\usepackage[usenames,dvipsnames]{xcolor}

\usepackage{amstext}
\usepackage{mathtools}
\usepackage{siunitx}
\usepackage{physics}
\usepackage{xspace}
\usepackage{multirow}
\usepackage{makecell}
\usepackage{booktabs}

\usepackage{tabularx}

\usepackage{newtxtext,newtxmath}

\usepackage{cleveref}

\makeatletter
\pretocmd\@sect{\def\@currentcounter{#1}}{}{\fail}
\makeatother

\DeclareSIUnit{\pc}{pc}
\DeclareSIUnit{\year}{yr}
\DeclareSIUnit{\Gyr}{Gyr}
\DeclareSIUnit{\Myr}{Myr}
\DeclareSIUnit{\dex}{dex}
\newcommand*{\Msun}{\ensuremath{\mathrm{M}_{\odot}}\xspace}

\newcommand*{\limepy}{\textsc{limepy}\xspace}
\newcommand*{\emacss}{\textsc{EMACSS}\xspace}
\newcommand*{\GCfit}{\textsc{GCfit}\xspace}
\newcommand*{\ssptools}{\textsc{SSPtools}\xspace}
\newcommand*{\clusterBH}{\textsc{clusterBH}\xspace}
\newcommand*{\cBHBd}{\textsc{cBHBd}\xspace}
\newcommand*{\BHBdynamics}{\textsc{BHBdynamics}\xspace}
\newcommand*{\Nbody}{\(N\)-body\xspace}
\newcommand*{\coupled}{\texttt{cBH+limepy}\xspace}
\newcommand*{\BSE}{\texttt{BSE}\xspace}
\newcommand*{\SSE}{\texttt{SSE}\xspace}
\newcommand*{\NBODYVII}{\texttt{NBODY7}\xspace}
\newcommand*{\PARSEC}{\texttt{PARSEC}\xspace}

\newcommand*{\uSSE}{u\texttt{SSE}\xspace}
\newcommand*{\uSSErapid}{u\texttt{SSE}-rapid\xspace}
\newcommand*{\uSSEdelay}{u\texttt{SSE}-delayed\xspace}
\newcommand*{\SEVN}{\texttt{SEVN}\xspace}
\newcommand*{\SEVNrapid}{\texttt{SEVN}-rapid\xspace}
\newcommand*{\SEVNdelay}{\texttt{SEVN}-delayed\xspace}

\newcommand*{\fkickeighty}{\ensuremath{f_{\mathrm{k}} = 80\%}\xspace}

\newcommand*\NGC[1]{NGC\thinspace{#1}}

\newcommand*{\omegacen}{\(\omega\)\thinspace{Cen}\xspace}

\newcommand*\chem[1]{\ensuremath{\mathrm{#1}}}
\newcommand*{\FeH}{\ensuremath{[\chem{Fe}/\chem{H}]}}

\newcommand*{\rhohi}{\ensuremath{\rho_{\mathrm{h},0}}\xspace}
\newcommand*{\Mi}{\ensuremath{M_{0}}\xspace}

\newcommand*{\fbhi}{\ensuremath{f_{\mathrm{BH},0}}\xspace}
\newcommand*{\fbh}{\ensuremath{f_{\mathrm{BH}}}\xspace}

\newcommand*{\Nbh}{\ensuremath{N_{\mathrm{BH}}}\xspace}
\newcommand*{\mbh}{\ensuremath{m_{\mathrm{BH}}}\xspace}

\newcommand*{\fret}{\ensuremath{f_{\mathrm{ret}}}\xspace}
\newcommand*{\fb}{\ensuremath{f_{\mathrm{b}}}\xspace}

\newcommand*{\vesci}{\ensuremath{v_{\mathrm{esc},0}}\xspace}

\newcommand*{\fkick}{\ensuremath{f_{\mathrm{k}}}\xspace}
\newcommand*{\RGeff}{\ensuremath{R^\prime_{\mathrm{G}}}\xspace}

\newcommand*{\rh}{\ensuremath{r_{\mathrm{h}}}}

\newcommand*{\Gaia}{\textit{Gaia}\xspace}
\newcommand*{\HST}{\textit{HST}\xspace}

\defcitealias{Fronimos2026}{F26}
\newcommand*{\cbhpaper}{\citetalias{Fronimos2026}\xspace}

\defcitealias{Dickson2026a}{Paper I}
\newcommand*{\paperI}{\citetalias{Dickson2026a}\xspace}

\begin{document}


\title{
    Fast Dynamical Modelling of Milky Way Globular Clusters --
    II. Impacts of Black Hole Prescriptions
}

\shorttitle{
    Fast Dynamical Modelling of MWGCs - BH Physics
}


\author[orcid=0000-0002-6865-2369,gname=Nolan,sname=Dickson]{Nolan Dickson}
\affiliation{Department of Astronomy and Physics, Saint Mary’s University, 923 Robie Street, Halifax, NS B3H 3C3, Canada}
\email[show]{nolan.dickson@smu.ca}

\author[orcid=0000-0003-2927-5465,gname=H\'enault-Brunet,sname=Vincent]{Vincent H\'enault-Brunet}
\affiliation{Department of Astronomy and Physics, Saint Mary’s University, 923 Robie Street, Halifax, NS B3H 3C3, Canada}
\email{vincent.henault@smu.ca}

\author[orcid=0000-0003-4158-5044,gname='Fronimos Pouliasis',sname=Fotios]{Fotios Fronimos Pouliasis}
\affiliation{Institut de Ciències del Cosmos (ICCUB), Universitat de Barcelona (UB), c. Martí i Franqués, 1, 08028 Barcelona, Spain}
\email{fotisfronimos@icc.ub.edu}

\author[orcid=0000-0002-9716-1868,gname=Gieles,sname=Mark]{Mark Gieles}
\affiliation{Institut de Ciències del Cosmos (ICCUB), Universitat de Barcelona (UB), c. Martí i Franqués, 1, 08028 Barcelona, Spain}
\affiliation{ICREA, Pg. Llu\'{i}s Companys 23, E08010 Barcelona, Spain}
\affiliation{Institut d'Estudis Espacials de Catalunya (IEEC), Edifici RDIT, Campus UPC, 08860 Castelldefels (Barcelona), Spain}
\email{mgieles@icc.ub.edu}

\author[orcid=0000-0001-6482-1842, sname=Daniel, gname='Marín Pina']{Daniel Marín Pina}
\affiliation{Zentrum für Astronomie der Universität Heidelberg, Institut für Theoretische Astrophysik, Albert-Ueberle-Str. 2, 69120 Heidelberg, Germany}
\email{daniel.marin@uni-heidelberg.de}

\author[orcid=0000-0002-7489-5244,gname=Smith,sname=Peter]{Peter J. Smith}
\affiliation{Max Planck Institute for Astronomy, K\"onigstuhl 17, D-69117 Heidelberg, Germany}
\affiliation{Department of Physics and Astronomy, University of Heidelberg, Im Neuenheimer Feld 226, D-69120 Heidelberg, Germany}
\email{pesmith@mpia.de}


\date{Accepted XXX. Received YYY; in original form ZZZ}



\begin{abstract}

    The populations of stellar-mass black holes (BHs) in globular clusters
    (GCs) play a key role in their dynamical evolution, however the mechanisms
    surrounding their formation and retention are uncertain.
    In this work, we extend the analysis of \paperI by fitting
    coupled rapid cluster evolution and multimass equilibrium models to
    a large sample of Milky Way GCs, under a variety of prescriptions
    for stellar evolution, BH formation and supernovae (SN)
    natal kicks.
    We explore the impacts of adopting \SSE or \PARSEC (through \SEVN)
    prescriptions for BH initial-final mass relations, the rapid or delayed SN
    fallback mechanisms, and an ad hoc grid of kick strengths
    ejecting between 40 and 80 per cent of all BHs formed.
    All models reproduce the same present-day conditions
    despite starting from notably different initial BH populations, due to the
    correlation found between the initial cluster densities and initial
    BH mass fractions.
    A linear relationship is found between the (log) initial half-mass density
    and the initial BH mass fraction, with the \SEVN models resulting in
    median densities (\(\rhohi \sim 10^{7.2\pm1.1}\,\unit{\Msun \ \pc^{-3}}\))
    nearly an order of magnitude higher than those of \SSE
    (\(\rhohi \sim 10^{6.4\pm0.9}\,\unit{\Msun \ \pc^{-3}}\)).
    We also find that both the bottom-light initial mass functions and
    the present-day BH mass fractions previously inferred
    are relatively robust against the stellar evolution models and natal kick
    prescriptions assumed.
    Finally, we discuss the implications of these results on the expected
    numbers and properties of dynamical binary-BH mergers, and the growth of
    intermediate-mass BHs.

\end{abstract}




\section{Introduction}\label{sec:introduction}


    The initial properties of the globular star clusters (GCs) we observe today
    in the local Universe are concealed by nearly a Hubble time of dynamical
    evolution.
    Yet these origins and early conditions are key for understanding, modelling,
    and contextualizing observations of a number of important processes such as
    star formation, massive black hole (BH) growth and binary-BH (BBH) mergers,
    as the initial conditions of these clusters determine their evolution
    over the billions of subsequent years.

    It is now understood that the populations of BHs within GCs play an
    important role in cluster evolution, as they segregate quickly to the
    core and generate the energy demanded by the rest of the cluster,
    through the formation and hardening of BBHs (in a process known as
    BH-burning), for much of its lifetime \citep{Henon1975,Breen2013b}.
    The number of BHs initially formed and retained in a cluster
    thus has a large impact on the expected mass loss and expansion rate of
    the system as a whole.
    However, the implementation in cluster simulations of these relationships
    between BHs and their host GC, ultimately relies on a number of
    assumptions, which introduce significant uncertainties.

    When the most massive stars born from a given IMF reach the ends
    of their lives and their cores collapse,
    they will often experience a supernova explosion (SN), and from the
    remaining material form either a Neutron Star (NS) or BH.
    The final mass of the newly formed remnant depends on both the mass of the
    star before collapse, primarily dictated by the stellar winds
    over the course of its lifetime, and on the amount of material ejected
    during the SN.
    In most cases, it is expected that asymmetries in these SNe will also
    impart a significant velocity to the newly formed remnants
    \citep{Janka1994,Burrows1996}.
    However, the exact mechanisms surrounding both these explosions and
    the preceding massive star evolution remain very
    uncertain, and depend on the complex physics underlying the
    structure and eventual collapse of the progenitor stars
    \citep{Belczynski2010,Burrows2021}.
    This is further complicated by the fact that most massive stars will form
    in binaries \citep[e.g.][]{Sana2013,Sana2025,Offner2023}, where additional
    binary processes will take place, and
    which also remain uncertain \citep[see e.g.][]{Guerrero2026}.
    In-depth modelling of the deaths of massive stars and their accompanying
    SNe is a difficult and computationally expensive task
    \citep[e.g.][]{Chan2018,Chan2020,Janka2024}, while direct
    (through e.g. the proper motions of isolated pulsars) and indirect
    (through e.g. remnant retention in star clusters) observations of the
    impacts of SNe kicks on their newly formed remnants do not yet provide a
    conclusive picture of the mechanisms and strengths of these natal kicks
    \citep[e.g.][]{Hobbs2005,Mandel2016,Repetto2017,Atri2019,Popov2025,
    Willcox2025a,Willcox2025b,Disberg2025}.

    Dynamical modelling and population synthesis studies typically include
    natal kicks by assuming a population-level
    distribution of kick velocities, such as a Maxwellian
    \citep[e.g.][]{Hobbs2005} or log-normal \citep[e.g.][]{Disberg2025} based
    on observational constraints. It is also common, due to the difficulties in
    directly measuring BH kicks, to assume that BHs simply follow a scaled
    version of the kick distributions applied to NSs \citep[e.g.][]{Fryer2012}.
    However, all of these approaches are plagued by the same uncertain
    foundations, may be calibrated on inconclusive observational evidence and
    are dependent on the underlying models of stellar evolution, which
    come with their own assumptions.

    While modern simulations of GC formation
    are able to simulate the births of clusters across a
    variety of environments \citep[e.g.][]{Cournoyer-Cloutier2024,Polak2024,
    Reina-Campos2025,Lahen2025a,Lahen2025b,Williams2025}, connecting their
    results with the GCs observed today is
    difficult, due to the aforementioned dynamical evolution.
    Direct star-by-star models, such as \texttt{NBODY} \citep{Aarseth2012},
    \texttt{PeTar} \citep{Wang2020}, \texttt{CMC} \citep{Kremer2020,
    Rodriguez2022} and \texttt{MOCCA} \citep{Hypki2013,Giersz2013}, have been
    used to explore the lifetimes of star clusters.
    However, these models come with significant computational costs, which
    scale with the number of objects.
    A less expensive approach is offered instead by rapid, semi-analytical
    models, which describe the bulk evolution of a cluster through simple
    prescriptions for the most relevant physical processes, rather than
    tracking the dynamics of individual objects.
    Examples include \emacss \citep{Alexander2012,Gieles2014,
    Alexander2014}, \textsc{Rapster} \citep{Kritos2024},
    \textsc{FastCluster} \citep{Mapelli2021} and \clusterBH
    \citep{Antonini2020a}, among others.
    Most relevant for this work, the \clusterBH models,
    which describe the co-evolution of both GCs and their hosted BH subsystems,
    were recently updated and expanded in \citet[][\cbhpaper]{Fronimos2026} to
    account for important processes such as different metallicities, variable
    initial stellar mass functions (IMFs), and the effects of galactic tidal
    fields.



    In \citet[][hereafter \paperI]{Dickson2026a}, we introduced a fitting
    pipeline which combined the fast evolutionary models of \clusterBH with
    multimass equilibrium models (\limepy), and used it to fit to a large
    sample of well-studied Milky Way (MW) GCs, in order to explore their
    initial conditions.
    We showed that the observational data favours
    clusters with relatively high initial densities, and that these could be
    degenerate with the initial
    populations of BHs formed and retained within the clusters.
    In this work, we explore this hypothesis and its implications in more
    detail, by revisiting the models and fitting framework of \paperI, and
    testing the impacts that various assumptions surrounding BH natal
    kicks and IFMRs may have on the inferred initial conditions.


    In \Cref{sec:coupled_models}, we briefly reintroduce the coupled
    \coupled models and pipeline we use to fit the clusters, as described in
    \paperI.
    In \Cref{sec:bh_methods}, we then describe the various prescriptions
    related to BH formation that we explore in this paper.
    The results of fitting our full sample of MW GCs under these
    prescriptions are then shown in \Cref{sec:results}, with a discussion of
    their impacts on the inferred initial conditions, IMFs and present-day BH
    populations in these clusters.
    Finally, in \Cref{sec:discussion}, the implications of these results for
    BBH mergers and IMBH formation are discussed, before we conclude in
    \Cref{sec:conclusions}.



\section{\coupled Models}
\label{sec:coupled_models}


    Modelling of the evolution of the bulk quantities of the clusters
    (e.g. total cluster mass \(M\), half-mass radius \rh, BH mass fraction \fbh)
    from their initial conditions to their present estimated age is undertaken
    by the \clusterBH fast evolutionary models
    \footnote{Available at \url{https://github.com/cBHBd/cBHBd}.}.
    \clusterBH was introduced by \citet{Antonini2020a,Antonini2020b,
    Antonini2023}, and recently updated in \cbhpaper to account for
    critical physical processes such as evaporation in an external
    galactic potential, different metallicities, and variable IMFs.
    In short, \clusterBH evolves a cluster from a set of initial conditions
    by accounting for the rates of change of the overall energy, total mass,
    half-mass radius and mass in BHs within the system dictated by a handful
    of key physical mechanisms.
    In particular, this includes
    the two-body relaxation demands of the cluster stars, which
    govern the energy produced within the core through the process of BH-burning
    by BBHs \citep{Breen2013b},
    the mass lost through the process of stellar evolution, as winds and
    SN dominate the cluster's early lifetime,
    and finally, the mass lost to the host galactic potential, with
    preferentially low-mass stars escaping from the cluster outskirts over the
    tidal boundary of the system.
    For more details, see \cbhpaper and \paperI.
    These models were calibrated and compared with a large grid of Cluster
    Monte Carlo \citep[CMC;][]{Kremer2020,Rodriguez2022} and \Nbody
    \citep{Gieles2021} models, and are able to accurately reproduce the
    evolution of the bulk quantities seen in these more computationally
    expensive models to within 20-30 per cent, across a wide range of cluster
    parameters, in fractions of a second of runtime, independent of the number
    of stars.


    To model in more detail the present-day phase-space distributions of stars
    and stellar remnants within GCs, we use the \limepy multimass, distribution
    function-based models \citep{Gieles2015} implemented through the \GCfit
    library \footnote{Available at \url{https://github.com/mgieles/limepy} and
    \url{https://github.com/nmdickson/GCfit}.}, as described in
    \citet{Dickson2023,Dickson2024}, and with the updates presented in \paperI.
    These mass models consist of the sum of component distribution
    functions which describe the phase-space densities of stars and stellar
    remnants in relaxed systems such as GCs, allowing us to account
    for the full spectrum of masses spanning main-sequence stars, white dwarfs
    (WDs), NSs and (key for this paper) BHs.


    To populate the present-day mass function (PDMF) of stars
    and the populations of stellar remnants,
    we use the mass function evolution algorithms implemented in the \ssptools
    library\footnote{Available at \url{https://github.com/SMU-clusters/ssptools}.}.
    These algorithms were first presented in \citet{Balbinot2018}, and then
    updated and expanded upon in \citet{Dickson2023} and \paperI.
    Starting from a stellar IMF (parametrized as a three-component
    power law with slopes \(\alpha_1,\,\alpha_2,\,\alpha_3\)), this model
    simulates the changes in the
    mass function over the lifetime of a cluster from both stellar evolution,
    as stars evolve off the main sequence and transform into stellar remnants,
    and through the loss of objects from the cluster via tidal evaporation,
    SNe natal kicks, and dynamical ejections.

    As WDs, NSs or BHs form, the types and masses of these remnants
    are determined based on their progenitor mass, metallicity and
    an initial-final mass relation (IFMR).
    Some fraction of the newly formed NSs and BHs receive natal kicks
    above the central escape velocity of the
    host cluster. To account for this, we scale the final numbers and total
    masses of remnants formed (and retained) within the cluster by an
    ``initial retention fraction'' \fret.

    We assume a constant NS retention fraction of \(\fret = 10\)
    per cent, as is common \citep[e.g.][]{Pfahl2002}
    \footnote{Because NS and BH formation are closely linked, many
    of the prescriptions we explore in this work would also affect the
    formation and retention of NSs; however it has been shown that, due to the
    low total mass in NSs, the results of mass modelling are insensitive to
    these effects \citep{Henault-Brunet2020}, and we thus do not explore them
    further here.}.
    The BH IFMR and natal kicks (the principal subjects of this paper) are
    described in more detail in \Cref{sec:bh_methods}.


    These models are combined into a single pipeline, which we
    hereafter refer to as \coupled.
    By first evolving the bulk quantities of the cluster from their initial
    conditions to the current cluster age (through \clusterBH),
    then determining the total masses and numbers of stars and remnants across
    an evolved mass function (through \ssptools), and finally computing the
    phase-space distribution of these objects within the cluster in equilibrium
    at the present day (through \limepy),
    we construct a forward model of GC evolution that can be compared
    directly with present-day observables to constrain initial conditions.

    In \paperI we utilized these \coupled models to fit a variety of
    observational datasets, across a sample of 40 MW GCs originally
    described in \citet{Dickson2023}, to infer the posterior
    probability distributions of the initial conditions and present-day model
    parameters that best describe each cluster.
    In this paper, we return to the same sample of clusters and, under a
    variety of different BH IFMRs and natal kick prescriptions, apply the
    same Bayesian parameter estimation methods (i.e. dynamic nested sampling;
    \citealt{Speagle2020}) using the same observational datasets, which
    includes proper motion (PM) dispersion profiles from \Gaia and \HST
    \citep{Watkins2015,Libralato2022,Haberle2025}, line-of-sight (LOS)
    dispersion profiles from various observatories and campaigns
    \citep{Lutzgendorf2013,Baumgardt2018,Kamann2018,Dalgleish2020},
    projected number density profiles from \Gaia and \HST
    \citep{Trager1995,Miocchi2013,deBoer2019} and present-day local
    stellar mass functions from \HST \citep{Baumgardt2023}.

\section{Black Hole Physics Prescriptions}
\label{sec:bh_methods}

    The principal objective of this manuscript is to extend the analysis of
    \paperI by exploring the relationship between our inferred initial
    cluster conditions and the prescriptions we adopt for the formation
    and retention of BHs.
    In particular, we investigate the impacts of different BH IFMR and
    natal kick prescriptions on the initial cluster
    masses, densities and stellar IMFs we infer.
    In this section, we describe each of the BH prescriptions that we
    apply and examine in this paper.

\subsection{updated-\SSE}\label{sub:usse-methods}

    The first approach, also used in \paperI, is to use the
    library of fitting formulae and recipes, \BSE/\SSE
    (Binary/Single Star Evolution)%
    \footnote{We will refer only to \SSE in this work, as we do not
    consider the impacts of binaries, in any of the approaches discussed here.
    Given that massive stars form with a high multiplicity fraction
    \citep[e.g.][]{Sana2013,Sana2025,Offner2023}, binary evolution
    could affect the initial populations of BHs, however these processes
    are also very uncertain \citep[see e.g.][]{Guerrero2026} and including them
    is beyond the scope of this work.}.
    The motivation behind this choice in \paperI was to match as closely as
    possible the prescriptions used in many common dynamical cluster models,
    such as \NBODYVII \citep{Aarseth2012}, MOCCA \citep{Hypki2013}, and the CMC
    models of \citet{Kremer2020b}, as we were
    validating our methods by fitting also to
    mock observations of the CMC public grid.
    In particular, we use the version of \SSE which was updated and presented
    by \citet[][hereafter referred to as \uSSE]{Banerjee2020}
    to include the new wind prescriptions of \citet{Belczynski2010},
    the pair-instability and pulsational pair-instability SNe
    mechanisms of \cite{Belczynski2016}, and the remnant
    formation and fallback prescriptions of \citet{Fryer2012}.

    Within the final phase of remnant formation from a SN, it is expected that
    material ejected by the explosion will form a shock front and transfer
    some of its energy to the surrounding gas.
    Under the conservation of momentum, some of this material may decelerate
    below the escape velocity of the newly formed proto-remnant and
    \textit{fall back} onto it, depositing additional mass \citep{Fryer2009}.
    The fallback fraction (\fb) refers to the fraction of the stellar envelope
    that falls back onto the BH. This fraction depends on the exact physics
    surrounding the SN engine mechanism,
    as well as the mass and metallicity of the progenitor star
    \citep{Fryer2012}.
    The presence of this fallback material plays a large role in determining the
    final expected BH masses, as well as the magnitude of natal kicks they
    receive, and is thus one of the main parametrizations we explore here.

    \citet{Fryer2012} introduced two different SN mechanisms, based on the
    timescale of the explosions.
    The first scenario, which was the setup used in \paperI and many other
    modelling efforts such as the CMC public grid, is the ``rapid'' SN
    scheme, which assumes the explosion occurs within 250 ms after the
    building neutron degeneracy pressure halts the stellar collapse and
    rebounds.
    In the second scenario, known as the ``delayed'' SN scheme, the explosions
    may take place over a longer timescale, which typically results in a
    weaker explosion.
    We hereafter refer to these two scenarios, within the \SSE models,
    as \uSSErapid and \uSSEdelay.
    These two methods differ mainly at lower masses; for progenitors with
    initial (zero age main sequence; ZAMS) masses \(\gtrsim \SI{30}{\Msun}\),
    the difference between the two explosion models is smaller.


    We use these models (and all setups discussed in this section)
    to dictate the formation processes of BHs, first through the BH IFMR, and
    then through the BH natal kicks.

    To determine the final masses of our newly formed BHs, based on
    the progenitor ZAMS masses drawn from an IMF, we must
    construct a functional form of the BH IFMR.
    To do so, we compute a grid of \uSSE models of single stars with
    ZAMS masses spanning \(m_{0}=\left[0,\,250\right]\,\Msun\) and
    metallicities spanning \(\FeH=\left[-2.50,\,+0.5\right]\), for both the
    \uSSErapid and \uSSEdelay setups.
    As we evolve the MF of a cluster with a given metallicity, we then
    interpolate the IFMR curve (as well as the minimum progenitor
    mass which will form a BH) from this grid, and assign newly
    formed remnants to the appropriate BH mass bin at each timestep.
    The exact parameters of the \uSSE models run for this grid are detailed in
    \Cref{table:sev_model_parameters}.


    As we form these BHs, we incorporate the subsequent BH
    natal kicks, by computing the fraction of these BHs that will be retained
    in the cluster after natal kicks (\fret).
    To do so, we first assume that the kick velocity is drawn from a
    Maxwellian distribution, with a dispersion of:
    \begin{equation}
        \label{eq:vdisp}
        \sigma(m_{0}) = (1-f_{\mathrm{b}}(m_{0}))\,\SI{265}{\km\per\s}
    \end{equation}
    where \(m_{0}\) is the initial (ZAMS) progenitor star mass and the
    base \SI{265}{\km\per\s} is the dispersion that has been found
    for NSs \citep{Hobbs2005}\footnote{It was found by \citet{Disberg2025}
    that an error in \citet{Hobbs2005} led to an overestimation of the scale
    of this velocity distribution, and the dispersion should be
    \(\sigma=\SI{215}{\km\per\s}\). However, this correction, while important
    for NS kicks, has a small impact on the overall retention of BHs, which
    is dominated more by the fallback fraction \fb. Therefore, for more direct
    comparison with many existing models like CMC, we use \SI{265}{\km\per\s}
    in this work.}, and which is then scaled down by the
    fallback fraction \fb, which as mentioned above is a function of the
    progenitor mass.
    We compute \fb in the same manner as we do the IFMRs, by interpolating
    from the grid of \uSSE models.

    We then compute \fret by integrating this Maxwellian velocity distribution
    up to the escape velocity \vesci:
    \begin{equation}
        \fret(m_0) = \operatorname{erf}
            \left( \frac{\vesci}{\sqrt{2}\sigma(m_{0})} \right)
            - \sqrt{ \frac{2}{\pi} }\,\frac{\vesci}{\sigma(m_{0})}\,
              \exp\left(\frac{-\vesci^{2}}{2\sigma^2(m_{0})}\right)
    \end{equation}
    where the dispersion \(\sigma(m_{0})\) is given by \Cref{eq:vdisp}, as a
    function of \fb and thus progenitor mass. We compute the initial escape
    velocity of the cluster based on the initial cluster mass and density, as
    done in \cbhpaper:
    \begin{equation}
        \vesci \simeq \SI{50}{\km\per\s}\,
            \left(\frac{\Mi}{{10^5}\,\unit{\Msun}}\right)^{1/3}
            \left(\frac{\rhohi}{{10^5}\,\unit{\Msun\pc^{-3}}}\right)^{1/6} .
    \end{equation}
    where the constant of proportionality holds for a \citet{King1966} model
    with \(W_0=7\).

\subsection{\SEVN}\label{sub:sevn-methods}

    The next set of prescriptions we examine use the rapid (binary) population
    synthesis library \SEVN
    \citep[Stellar EVolution for \Nbody; ][]{Spera2019,Iorio2023}.
    In contrast to many other rapid population synthesis models, \SEVN is not
    based on the fitting formulae and recipes of, e.g., \SSE, but instead
    determines the evolution of stars by interpolating their stellar properties
    along pre-computed stellar evolution tracks.
    This foundation allows for an extra degree of flexibility in the \SEVN
    models, as the desired stellar tracks can be changed easily.
    For this work, the \SEVN models allow for the exploration of the impacts of
    notably different stellar evolution prescriptions (and thus BH IFMRs) than
    many other available options (e.g. \texttt{MOBSE}, \citealt{Giacobbo2018};
    \texttt{COMPAS},  \citealp{Riley2022}; \texttt{COSMIC},
    \citealp{Breivik2020}), all of which differ only slightly from \uSSE
    since they are rooted in the same \SSE prescriptions of
    \citet{Hurley2000,Hurley2002}.

    In this work, we use the default lookup tables from \SEVN, which correspond
    to the latest \PARSEC
    stellar tracks \citep{Bressan2012,Tang2014,Chen2015}.
    The exact parameters of the \SEVN models used are detailed in
    \Cref{table:sev_model_parameters}.
    As in \Cref{sub:usse-methods}, we examine two sets of fits under these
    models, applying both the rapid (\SEVNrapid) and delayed (\SEVNdelay)
    SNe prescriptions of \citet{Fryer2012}.
    We then follow the same procedure as before, and generate a grid of
    \SEVN models across initial progenitor masses and metallicities, in order
    to extract the functional forms of the BH IFMRs and \fb.
    The natal kicks are then also applied in the same manner.

    While we denote these models as \SEVN, note that, as we
    do not employ any binary evolution prescriptions, \SEVN acts solely as an
    interpolator for the \PARSEC stellar evolution tracks, and thus our
    results reflect primarily the \PARSEC stellar evolution tracks.
    Employing different stellar evolution tracks within \SEVN, such as the
    latest MIST tracks \citep[v2.5;][]{Dotter2026,Bauer2026}, would lead to
    correspondingly different results.

\subsection{Simple Kick Models}

    Within the existing stellar evolution models and BH
    prescriptions, including those explored here, significant uncertainties
    remain.
    In particular, for BH natal kicks, extensive modelling of
    SN explosions and remnant formation remains difficult
    (see Section 3.5 of \citet{Fryer2012} for a discussion of the
    uncertainties and caveats that exist within their fallback
    mechanisms alone) and none of the currently available rapid prescriptions
    are able to capture the full complexity of SN physics
    \citep[e.g.][]{Burrows2019,Vartanyan2019}.
    As such, we aim to detach our analysis from the choice of a specific
    setup of BH physics, and instead examine the overall impacts which BH natal
    kick distributions of various strengths might have on the inferred initial
    conditions of our clusters.
    Therefore we also introduce a simple parametrization of the natal kick
    retention fraction (\fret) that is independent of any physical
    cluster quantities, and that effectively sets the initial
    retained BH mass fraction \fbhi.

    To motivate this parametrization, we note that each
    of the other physically-motivated prescriptions described above share
    a common \fret shape; lower-mass progenitors generally receive stronger
    kicks, while higher-mass progenitors, which form their BHs through direct
    collapse without a SN, receive no kicks at all. Between these regimes there
    is a smooth turnover, typically near \(m_0\sim\SI{30}{\Msun}\).
    Therefore, we introduce a new sigmoid retention function:
    \begin{equation}\label{eq:fret_sigmoid}
        \fret(m_0) = \frac{1}{2} \left(
            \tanh\left(\mathrm{u_{\mathrm{k}}}\ (m_0 - m_{\mathrm{k}})\right) + 1
        \right).
    \end{equation}
    This function increases smoothly from 0 to 1, and is characterized
    by the scale mass \(m_{\mathrm{k}}\), marking the transition point between
    0\% and 100\% retention (such that \(\fret(m_{\mathrm{k}})=50\%\)), and the
    slope parameter \(u_{\mathrm{k}}\), which sets the steepness of this
    turnover.

    It is more useful, however, to parametrize this function not in terms
    of \(m_{\mathrm{k}}\), but in terms of the total fraction of BHs ejected,
    in order to describe impacts of the ``strength'' of the kicks.
    Therefore, we also introduce the parameter \fkick, bounded between 0 and 1,
    which is defined as the total fraction of initially formed BH mass
    ejected from the cluster due to natal kicks, such that:
    \begin{equation}
        \fkick = 1 - \frac{M_{\mathrm{BH},\mathrm{f}}}{M_{\mathrm{BH},0}}
    \end{equation}
    where \(M_{\mathrm{BH},0}\) is the total mass of BHs formed
    originally (from all stars that will form BHs), and
    \(M_{\mathrm{BH},\mathrm{f}}\) is the total mass of BHs retained
    in the cluster after natal kicks are applied using the retention function
    in \Cref{eq:fret_sigmoid}.
    Given some desired value of \fkick (and a slope \(u_{\mathrm{k}}\)), we
    employ a simple root finding algorithm to determine what scale mass
    \(m_{\mathrm{k}}\) will eject the
    desired total amount of BH mass. This lets us shift the
    characteristic mass of this retention function in order to reach a specific
    fraction of BH natal ejections, independent of the cluster
    metallicity or escape velocity.

    In this paper, we explore the impacts of a grid of \fkick values
    (\(\fkick=(40,\,60,\,80)\,\%\)).
    For these models, we use the BH IFMR from \uSSErapid.
    The slope \(u_{\mathrm{k}}\) is allowed to vary freely.

\subsection{Comparisons of BH Prescriptions}

    In \Cref{tab:prescriptions}, we present an overview of the seven different
    models that we fit to our entire sample of MW clusters, as described above.

    \begin{table}
        \centering
        \begin{tabular}{ c c c c }
            \hline
            Model Name      & BH IFMR        & Natal Kick & Fallback \\
                            & Method         & Method     & Method   \\
            \hline
            \uSSErapid      & \uSSE          & Maxwellian & rapid    \\
            \uSSEdelay      & \uSSE          & Maxwellian & delayed  \\
            \SEVNrapid      & \SEVN (\PARSEC) & Maxwellian & rapid    \\
            \SEVNdelay      & \SEVN (\PARSEC) & Maxwellian & delayed  \\
            \(\fkick=40\%\) & \uSSE          & Sigmoid    & rapid    \\
            \(\fkick=60\%\) & \uSSE          & Sigmoid    & rapid    \\
            \(\fkick=80\%\) & \uSSE          & Sigmoid    & rapid    \\
            \hline
        \end{tabular}
        \caption{
            List of the different models explored in this work and the
            natal kick, fallback and BH IFMR prescriptions employed in each.
        }
        \label{tab:prescriptions}
    \end{table}


    \begin{figure}
        \centering
        \includegraphics[width=\linewidth]{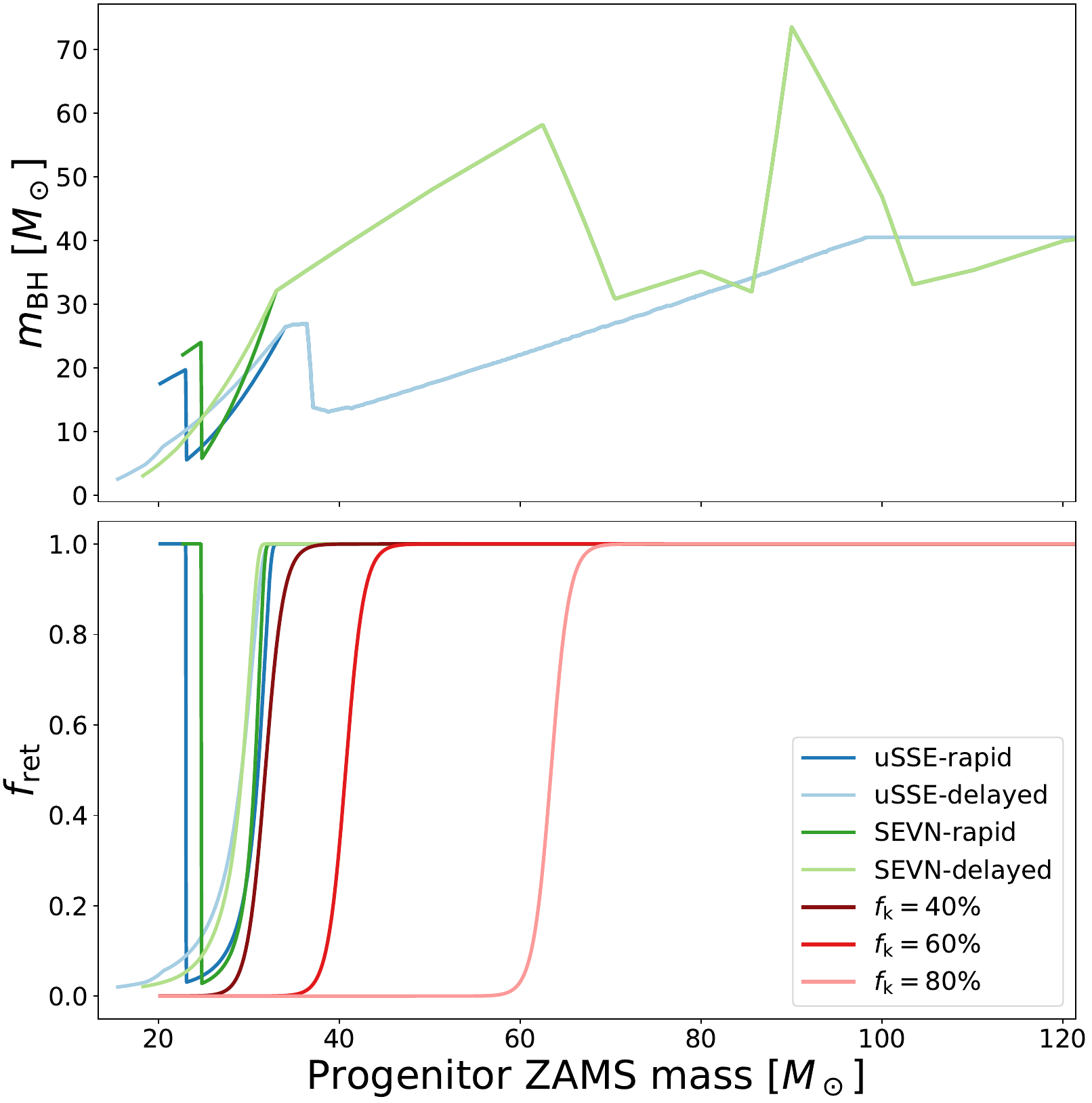}
        \caption{
            Example BH IFMRs (top panel) and BH kick retention functions \fret
            (bottom panel) for an example set of relevant parameters
            (\(\FeH=-1.5\), \(\vesci=\SI{100}{\km\per\s}\), \(\alpha_1=-0.82\)
            and \(\alpha_2=-1.45\)), under each BH prescription explored.
            The \fkick models use the BH IFMR of \uSSErapid, and are shown here
            with a \fret slope parameter of \(u_{\mathrm{k}}=0.5\).
        }
        \label{fig:ifmr_fret_methods}
    \end{figure}

    In \Cref{fig:ifmr_fret_methods}, we show the BH IFMRs and the retention
    functions for an example set of parameters (median IMF slopes from \paperI,
    \(\FeH=-1.5\), \(\vesci=\SI{100}{\km\per\s}\)) across all of the
    prescriptions that we explore in this paper.
    As mentioned in \Cref{sub:usse-methods}, it is clear from this that, for
    a given stellar evolution model, the rapid and delayed SN mechanisms vary
    only noticeably at lower-mass BH progenitors. Meanwhile, the IFMRs of the
    \uSSE and \SEVN models are quite different, with \SEVN having generally
    higher minimum BH-progenitor masses, but also creating more massive BHs
    in most cases.
    \Cref{fig:ifmr_fret_methods} also shows how the \fkick methods require
    increasingly larger \(m_{\mathrm{k}}\) values in order to reach the target
    kick fractions.
    Each of these \fkick values is chosen to provide stronger kicks
    than the physical models.
    Note that, by design, the \fkick methods remove the
    desired fraction of BHs independent of the rest of the cluster parameters,
    such as the cluster escape velocity. Meanwhile, the other methods
    do vary, kicking out between \(\sim5-35\,\%\) of their BHs over
    escape velocities ranging between \(\sim75-\SI{300}{\km\per\s}\) (as seen
    in the results of \paperI), with the \SEVN and delayed models kicking out
    more than the \uSSE or rapid models.


    \begin{figure}
        \centering
        \includegraphics[width=\linewidth]{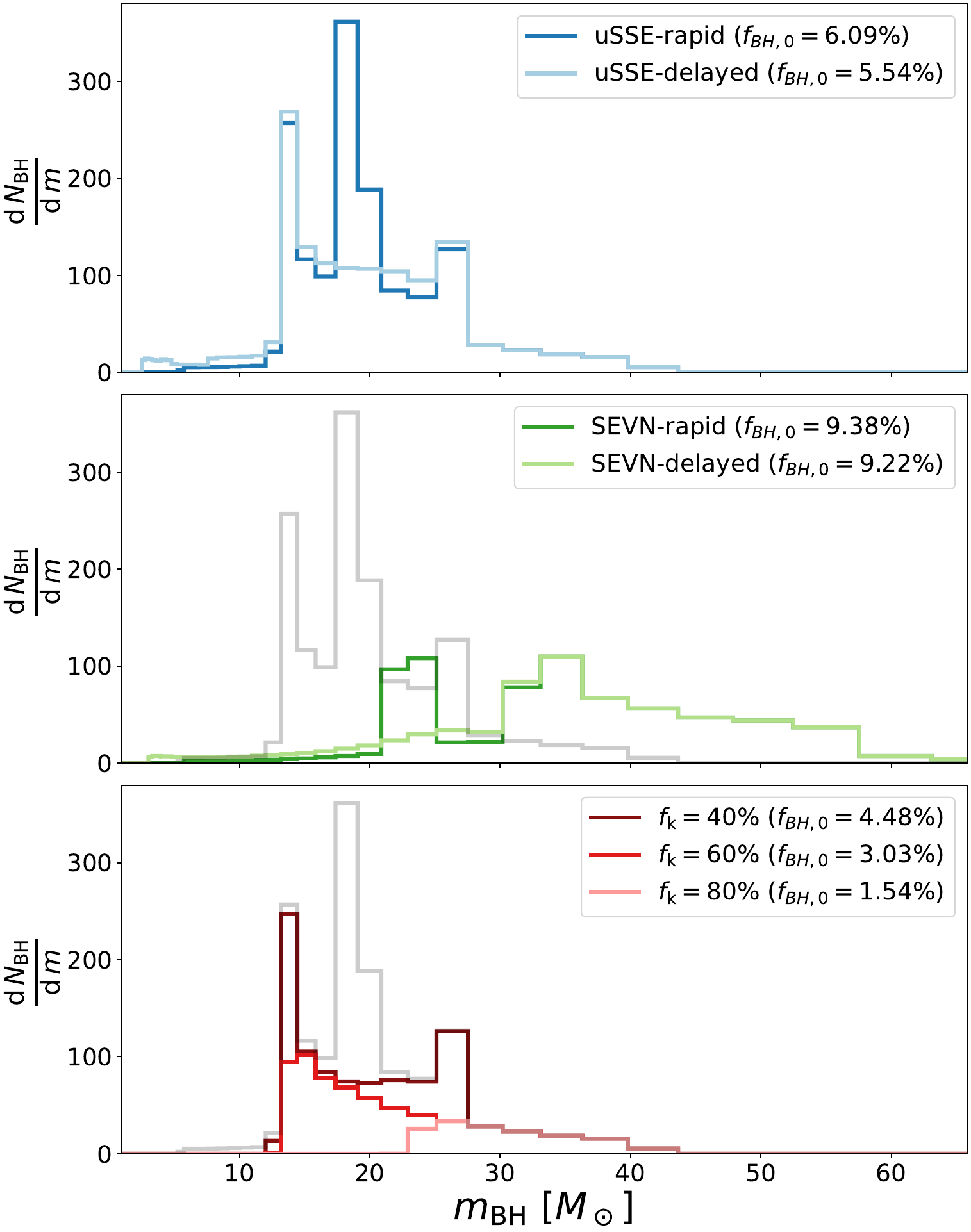}
        \caption{
            Example BH mass functions (including all BHs that will be formed
            and retained in the cluster after natal kicks) for an example set
            of relevant parameters (\(\FeH=-1.5\),
            \(\vesci=\SI{100}{\km\per\s}\), \(\alpha_1=-0.82\) and
            \(\alpha_2=-1.45\)), under the \uSSE (top panel),
            \SEVN (middle panel) and \fkick (bottom panel) BH prescriptions.
            The \uSSErapid BH mass function is shown in grey in all panels, for
            comparison.
            The corresponding initial BH mass fractions \fbhi are noted in the
            legend of each panel.
        }
        \label{fig:ibh_methods}
    \end{figure}

    In \Cref{fig:ibh_methods}, we show the total initial BH mass functions
    (for this same example model), which includes all BHs that will be formed
    and retained in the cluster after natal kicks, across all of our
    prescriptions.
    While the delayed SN mechanism shrinks the peaks of and generally broadens
    the BH distributions, relative to the rapid case, the overall differences
    in \fbhi are not typically large between the two
    (maximum of \(\lesssim 0.5\%\)).
    On the other hand, the \uSSE and \SEVN models are more noticeably different,
    with \uSSE peaking around \(\mbh\sim\SI{15}{\Msun}\) and \SEVN, with a
    distribution reaching much higher BH masses, peaking closer to
    \(\mbh\sim\SI{35}{\Msun}\).
    Consequently, and most importantly for our analysis, the \SEVN models
    have a much higher value of \fbhi, nearly 3 percentage points
    higher than the corresponding \uSSE models, for these example parameters.

    In this work we aim to examine a set of prescriptions which spans a wide
    range of \fbhi, rather than an exhaustive list of all available
    stellar evolution models, however most of the results of \Cref{sec:results}
    should be extensible to other models with different values of \fbhi.




\section{Results}
\label{sec:results}

    \begin{figure}
        \centering
        \includegraphics[width=\linewidth]{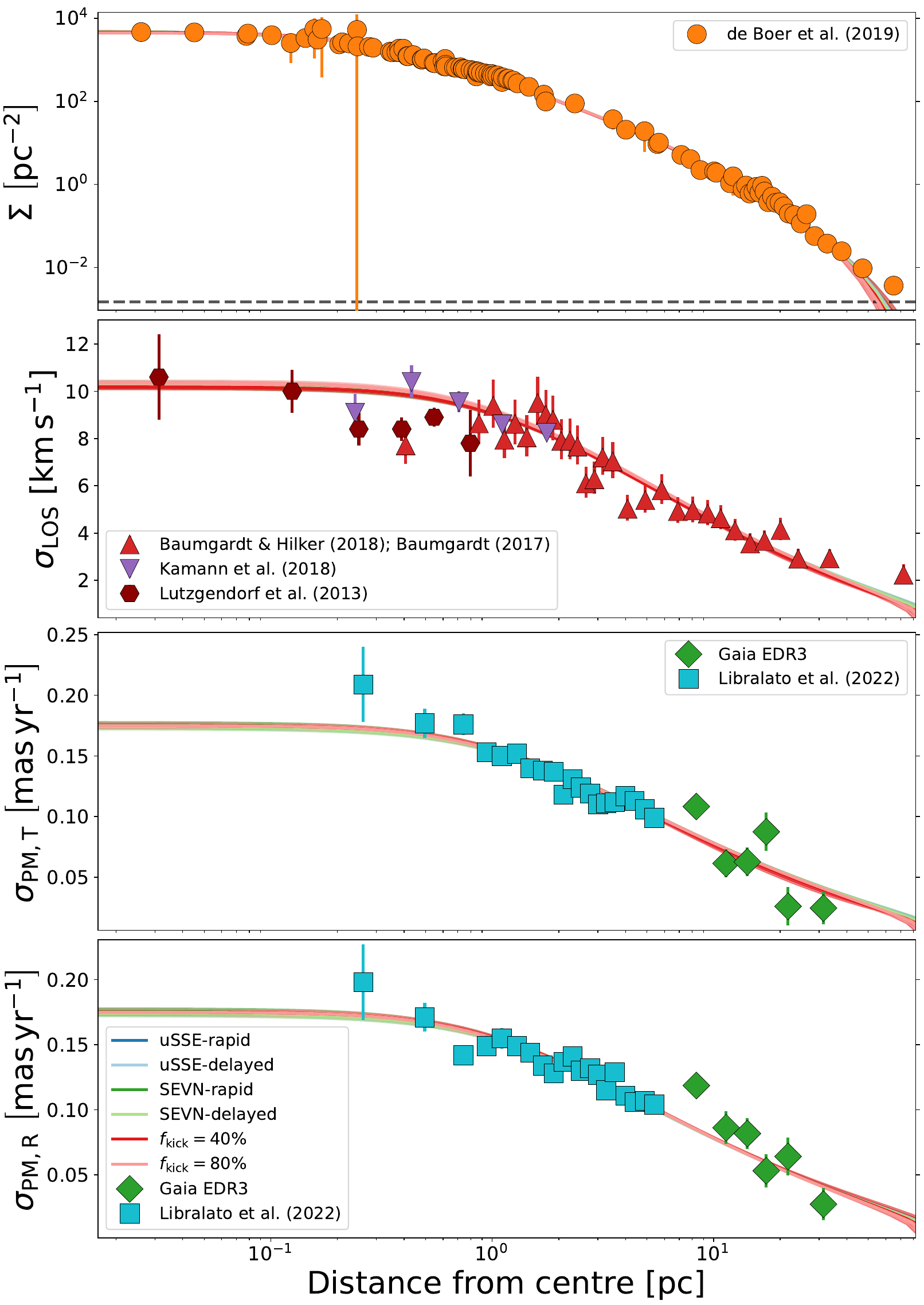}
        \caption{
            Model radial profiles (blue contours) of
            surface number density (\(\Sigma\)),
            line-of-sight velocity dispersions (\(\sigma_{\mathrm{LOS}}\)),
            radial (\(\sigma_{\mathrm{PM},\mathrm{R}}\)) and
            tangential (\(\sigma_{\mathrm{PM},\mathrm{T}}\)) proper motion
            dispersions, for the fit of \NGC{1851} under each BH
            prescription we explore.
            The dark and light shaded regions represent the \(1\sigma\)
            and \(2\sigma\) credible intervals of the model fits, respectively.
            The observational datasets used to constrain the models are shown
            alongside their \(1\sigma\) uncertainties by the various
            markers and errorbars.
            The background value subtracted from the number density profile is
            shown by the dashed line.
        }
        \label{fig:profile_fit}
    \end{figure}

    \begin{figure}
        \centering
        \includegraphics[width=\linewidth]{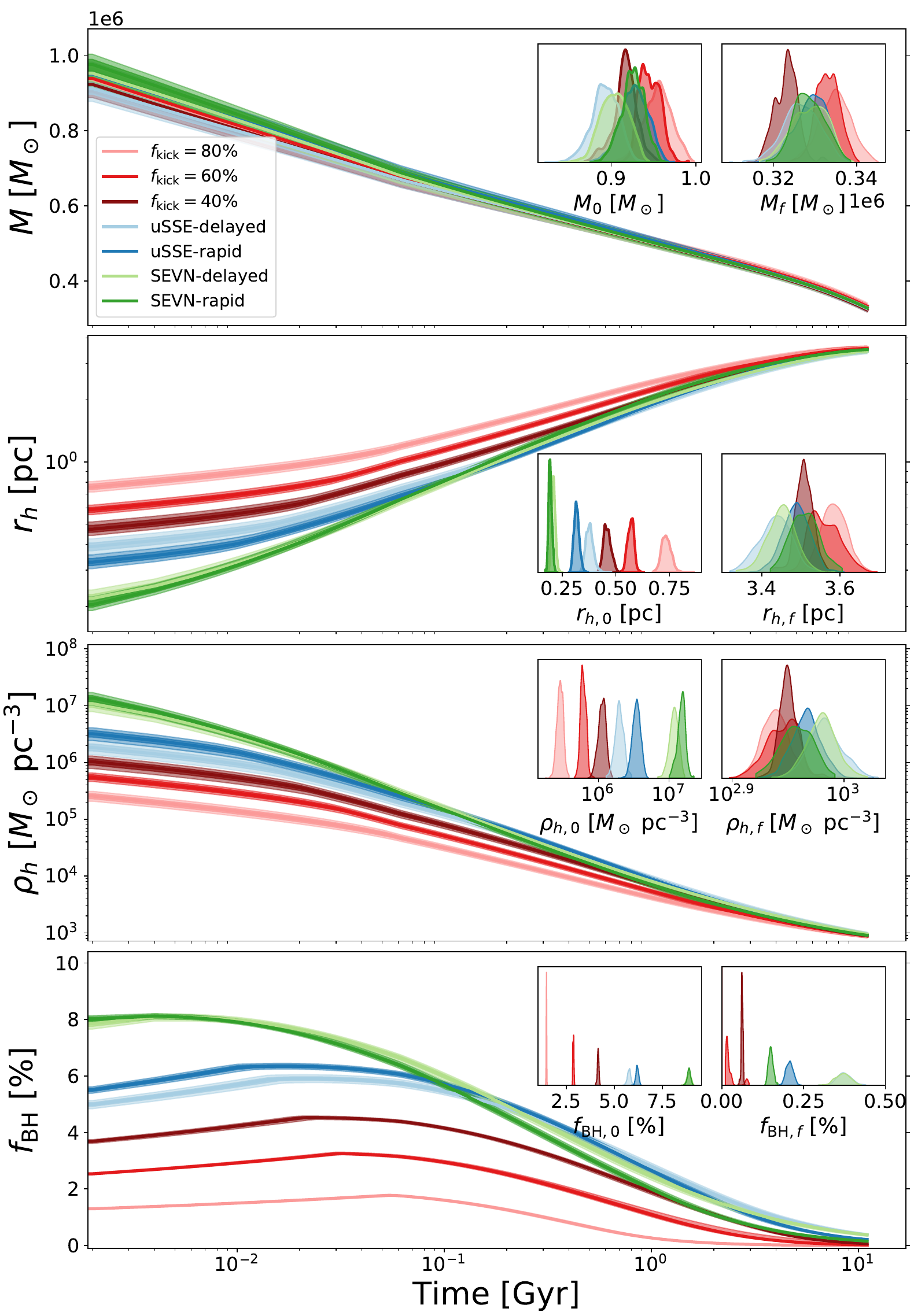}
        \caption{
            Evolution of cluster total mass, half-mass radius, half-mass density
            and BH mass fraction over time for the fits of \NGC{1851}, under
            each BH prescription we explore.
            The dark and light shaded regions represent the \(1\sigma\)
            and \(2\sigma\) credible intervals of the model fits,
            respectively.
            Within each panel, the full posterior distributions from our fits
            for both the initial and final values of each quantity are shown.
        }
        \label{fig:evolution_fit}
    \end{figure}

    In \Cref{fig:profile_fit}, we show an example fit of our models to
    some of the observational datasets of \NGC{1851}.
    As was shown in \paperI, the best-fitting models match the observational
    data well for the majority of clusters in our sample\footnote{Similar
    figures showing the fits to all models in the sample, as well as full
    sampler chains, can be found at \url{http://doi.org/10.11570/26.0017}.}.
    Notably, our present-day models under all BH prescriptions are
    nearly indistinguishable from one another.
    The reasons for this become apparent in \Cref{fig:evolution_fit}, where we
    show the inferred evolution of the mass, radius, density and \(\fbh\) from
    initial conditions to their present-day values.
    We can clearly see that the final, present-day properties
    of each model are very similar to one another.
    This is despite the fact that the initial conditions of nearly all
    properties (with the possible exception of the cluster mass) are very
    different.
    As expected, the initial BH mass fractions are successively lower for
    each setup with stronger kicks or IFMRs peaked at lower BH masses,
    following what was shown in \Cref{fig:ibh_methods}.
    Thus, to end up with the same conditions today, as shown
    in \Cref{fig:profile_fit}, the initial densities
    of these clusters are correspondingly lower, driven by an increase in
    the initial radii.
    The physical explanation for this is that the amount of ejected BH mass
    scales with the cluster density as
    \(\Delta M_{\mathrm{BH}}\propto \sqrt{\rho_{\mathrm{h}}}\)
    \citep{Breen2013b}.

    For the same reasons as discussed in \paperI (Section 4.1), the best-fitting
    models for a few clusters (\NGC{288}, \NGC{5927}, \NGC{6624}, \NGC{6656}
    and \NGC{6981}) do not adequately reproduce the observational datasets
    under most BH prescriptions.
    Therefore, we remove these from the original sample again, and proceed
    with the same final 35 clusters as in \paperI.
    We also note again that, as in \paperI, the fits to clusters which orbit at
    low effective galactocentric orbits (\(\RGeff\lesssim\SI{1.5}{\kilo\pc}\))
    and to core-collapsed clusters which retain significant BH populations
    should be regarded with caution. These are denoted by a dagger and asterisk,
    respectively, in the rest of this paper.

\subsection{Initial Conditions}
\label{sub:initial_conditions}

    \begin{figure*}
        \centering
        \includegraphics[width=\linewidth]{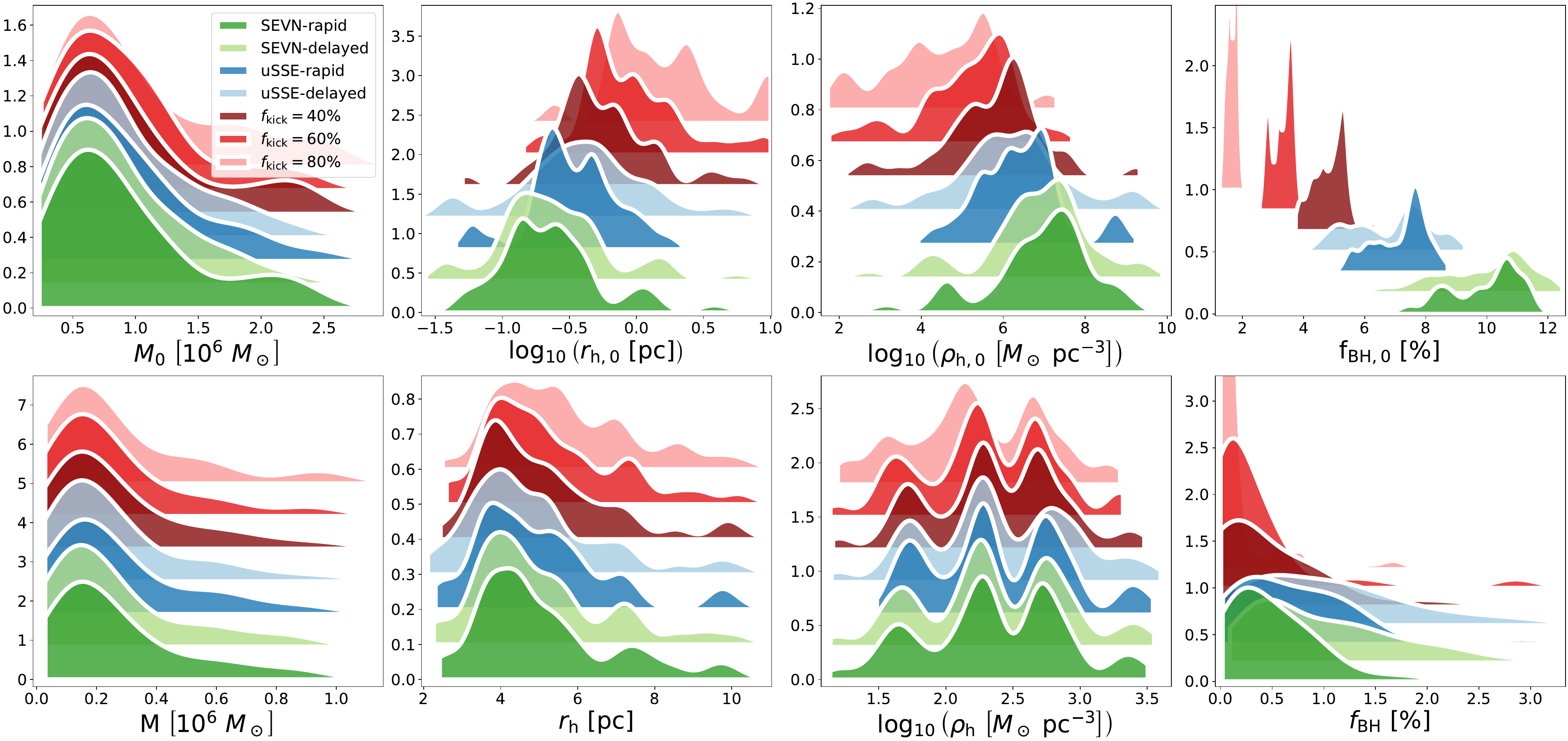}
        \caption{
            Gaussian kernel density estimates showing the distributions
            of the initial (top row) and final (bottom row) total mass,
            half-mass radius, half-mass density and BH mass fractions across
            the fits to all clusters in our sample, under each BH prescription
            explored.
            The distributions are shifted vertically to aid comparison.
        }
        \label{fig:param_dists}
    \end{figure*}

    In \Cref{fig:param_dists}, we show Gaussian kernel density estimates
    of the overall distributions of the initial and final
    mass, radius, density and BH mass fraction, across the fits to all clusters
    in our sample, for each of the BH prescriptions we explore.
    The present-day distributions of each of these quantities are similar
    across all prescriptions, in line with the example fits shown above.
    The inferred initial cluster masses are also similar across the different
    prescriptions. This indicates that the mass loss rates
    of these clusters, which are set on the same orbits in each method,
    are therefore broadly similar across BH prescriptions.
    The initial BH mass fractions vary across prescriptions, as is expected
    based on \Cref{sec:bh_methods}.
    Correspondingly, the distributions of initial radii are lowest under the
    prescriptions with higher \fbhi, resulting in higher initial densities.

    For many clusters, under the \SEVN prescriptions, we infer an initial
    half-mass radius which is even smaller than originally inferred in \paperI,
    typically less than \SI{0.5}{\pc} and peaking at around \SI{0.15}{\pc}
    (compared to \SI{0.7}{\pc} and \SI{0.25}{\pc}, respectively, for the \uSSE
    case).
    This in turn leads to a distribution of \rhohi that is shifted to higher
    values, with a median value of around \(10^{7.2}\, \unit{\Msun\ \pc^{-3}}\)
    and extending just past \(10^{8}\, \unit{\Msun \ \pc^{-3}}\).

    \begin{figure}
        \centering
        \includegraphics[width=\linewidth]{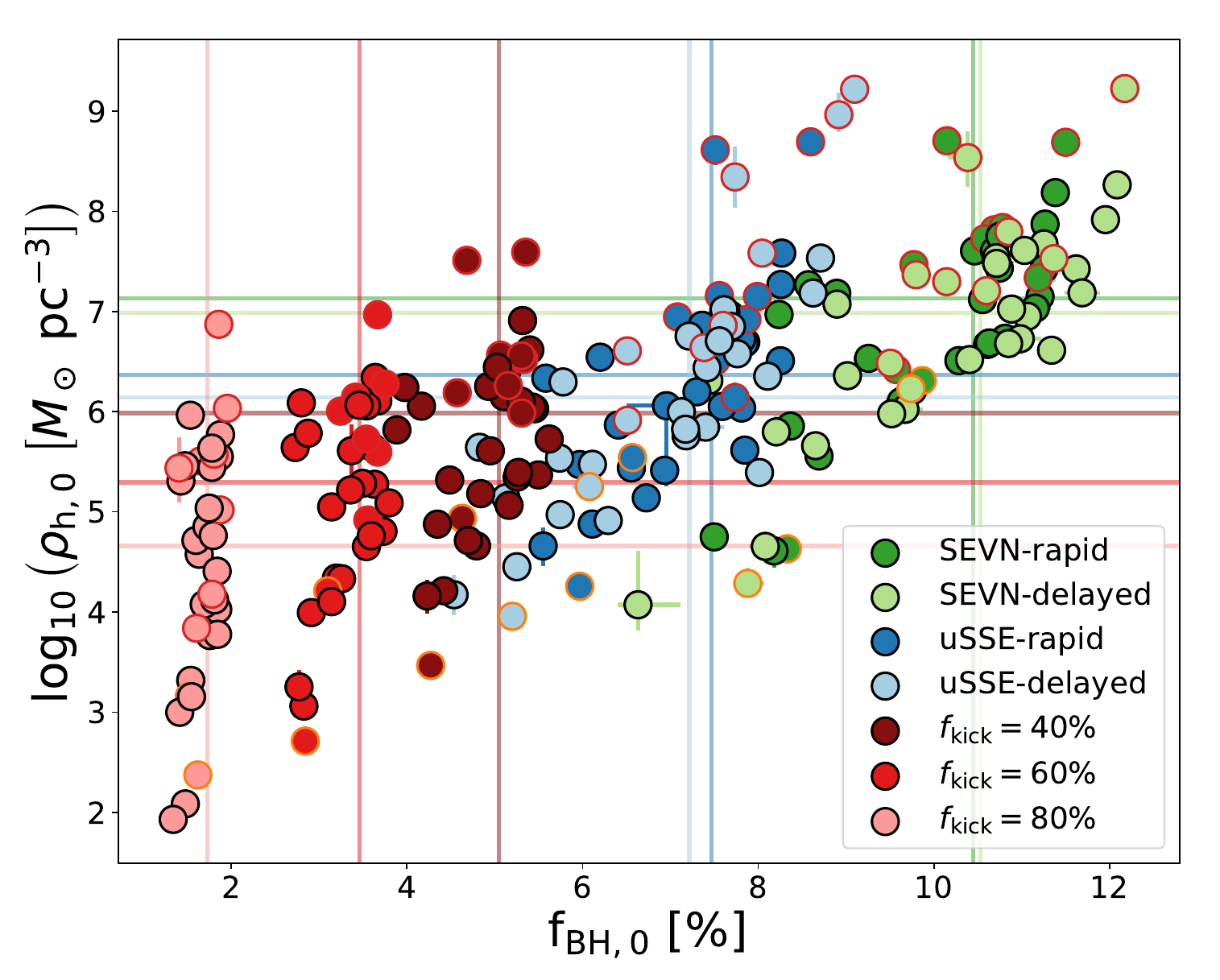}
        \caption{
            Relationship between the inferred initial BH mass fractions \fbhi
            and the initial half-mass (log) densities \rhohi across all
            clusters in our sample, across all BH prescriptions explored.
            The median values of \fbhi and \rhohi in each prescription are
            shown as vertical and horizontal lines.
            All core-collapsed clusters are outlined in red.
            All clusters with \(\RGeff<\SI{1.5}{\kilo\pc}\) are outlined in
            orange.
        }
        \label{fig:fbh0_vs_density}
    \end{figure}

    \begin{figure}
        \centering
        \includegraphics[width=\linewidth]{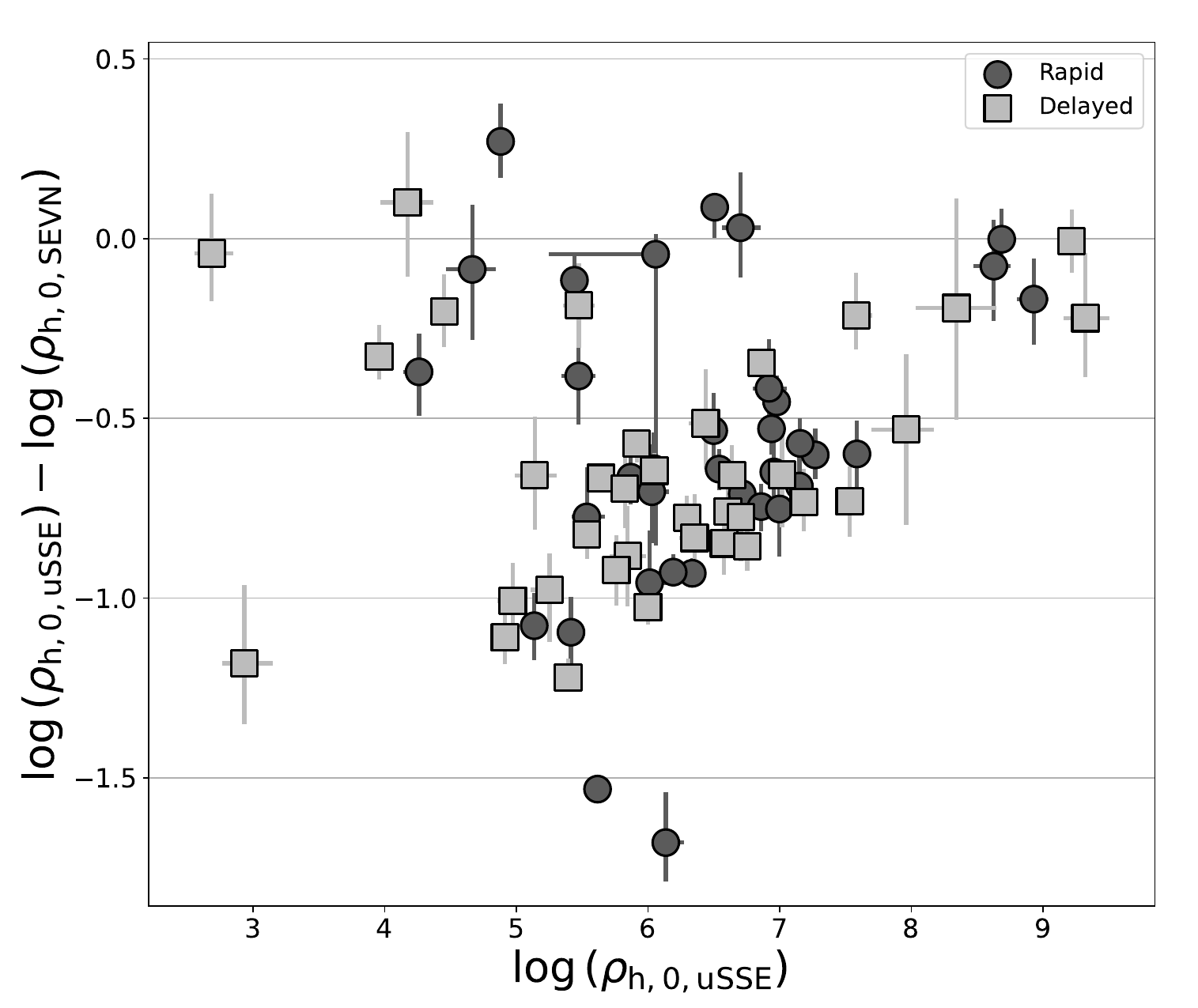}
        \caption{
            The residuals between the initial half-mass densities inferred
            under the \uSSE and \SEVN BH prescriptions, for all clusters in
            our sample.
        }
        \label{fig:density_residuals}
    \end{figure}

    In \Cref{fig:fbh0_vs_density}, we emphasize the differences in the initial
    density distributions between the different prescriptions, by plotting
    the inferred initial densities against the initial BH mass fractions
    (which captures the differences between methods from variations in both
    natal kick and IFMR prescriptions).

    Most notable here is the relationship between \fbhi and \rhohi.
    Across the various methods, as \fbhi increases, there is also a clear
    trend of higher inferred \rhohi.
    From the aforementioned theory of BH ejection \citep{Breen2013b} we would
    naively expect \(\rhohi \propto \fbhi^2\), however the scaling we find is
    steeper than this, likely because the clusters spend only a short time at
    higher densities. While the overall distributions of
    density overlap significantly, it is clear from the median values and
    distribution peaks that a systematic linear
    relationship exists between \fbhi and the log of \rhohi.
    The \SEVN models, for example, have a median value of initial densities
    shifted nearly 1 dex above the corresponding \uSSE models.
    This can be examined more clearly in \Cref{fig:density_residuals}, where we
    show the residual difference between the \uSSE and \SEVN methods, and can
    see that the majority of clusters are between 0.5 and 1 dex higher in the
    \SEVN case.

\subsection{Initial Mass Functions}
\label{sub:initial_mass_functions}

    \begin{figure}
        \centering
        \includegraphics[width=\linewidth]{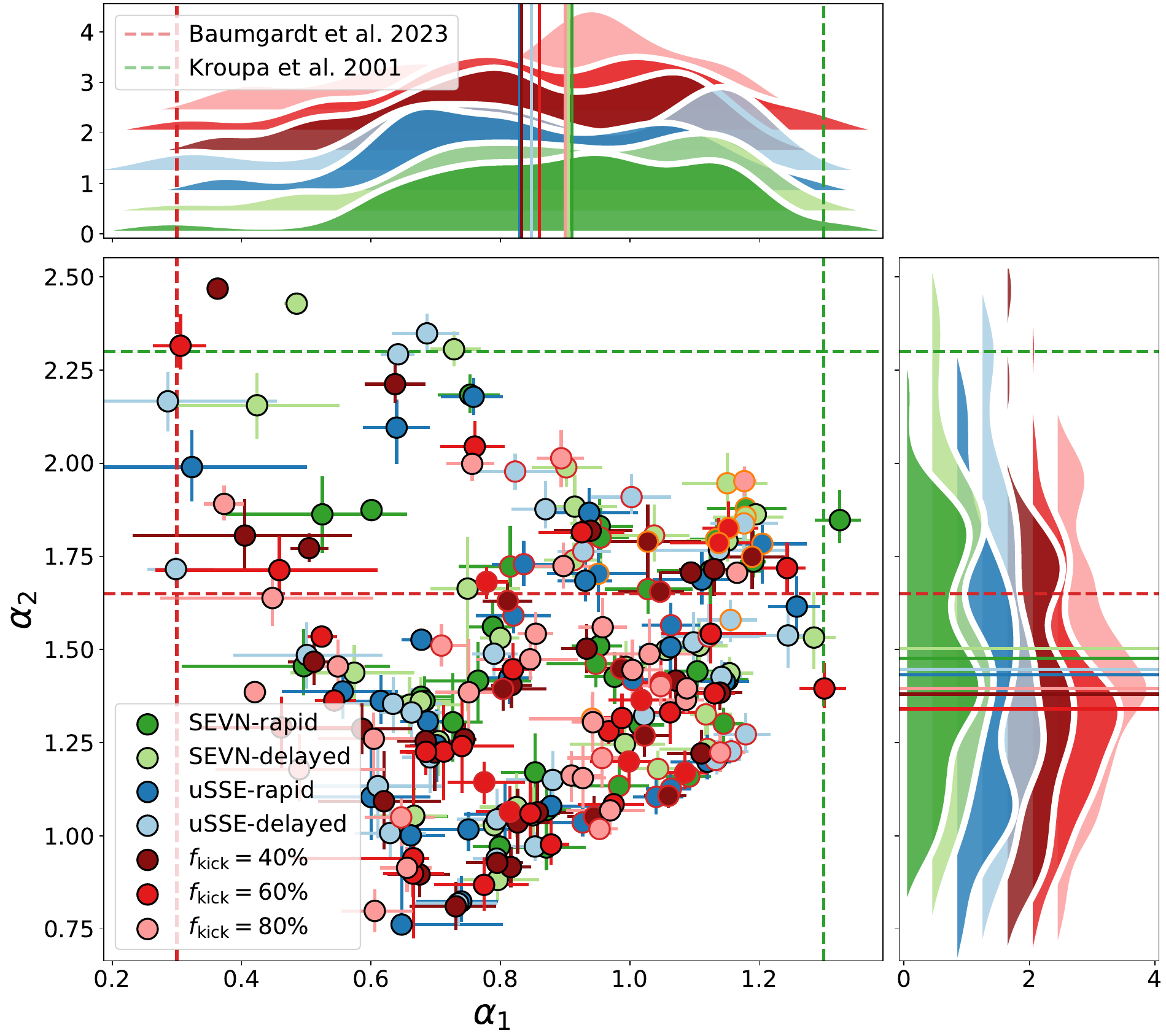}
        \caption{
            The median and \(1\sigma\) uncertainties of the inferred stellar IMF
            power-law slopes \(\alpha_1\) and \(\alpha_2\) for all clusters in
            our sample, under each BH prescription explored.
            The distributions of each parameter are also represented by
            a Gaussian kernel density estimate above and to the right, and the
            medians of each distribution are shown as vertical and horizontal
            lines.
            The distributions are shifted along their vertical axis to aid
            comparison.
            The median values of the corresponding IMF slopes from
            \citet{Kroupa2001} (green) and \citet{Baumgardt2023} (red) are also
            shown as the horizontal and vertical dashed lines on each panel.
            All core-collapsed clusters are bordered in red.
            All clusters with \(\RGeff<\SI{1.5}{\kilo\pc}\) are bordered in
            orange.
        }
        \label{fig:IMF_slopes}
    \end{figure}

    In \paperI we examined the values of the IMF power-law slopes below
    \SI{1}{\Msun} (\(\alpha_1,\,\alpha_2\)) and found, as in other recent works
    \citep{Dickson2023,Baumgardt2023}, that a bottom-light IMF, deficient in
    low-mass stars relative to a more canonical IMF like that of
    \citet{Kroupa2001}, was preferred by the data.
    This shallow IMF leads to significantly higher initial BH mass fractions,
    in comparison to canonical prescriptions, and is one of the main reasons
    that relatively high initial densities were found for most clusters in
    \paperI.

    In \Cref{fig:IMF_slopes} we show these same slopes, for each BH
    prescription that we examine.
    It is clear from this figure that, no matter the initial density or BH
    distribution, our fits consistently recover similar IMF slopes across all
    prescriptions.
    The medians of each prescription do not deviate by more than 0.1 from one
    another in either slope, and most individual clusters are consistent in
    their inferred IMF under each method.

    This demonstrates that the IMF slopes are most directly
    constrained by present-day MF observations, while accounting for the
    gradual change of the MF slopes, mostly from tidal evaporation, which is
    itself primarily a function of the cluster orbits and total mass.
    As shown above, the inferred cluster masses, over the entire evolution,
    do not change very much across our different prescriptions (with the
    differences in density being wrought by the differences in radii, not mass),
    and therefore we would expect the change in the MF to be similar
    across all prescriptions.
    This highlights that our inferred, non-canonical IMF is robust against
    changes to the stellar evolution and remnant prescriptions used, and
    driven by the data.

    Note that, as was discussed in \paperI, the inferred IMF slopes will be
    somewhat degenerate with the strength of the tidal field a cluster
    experiences over its lifetime. However, a much stronger tidal field would be
    required to bring our IMFs in line with canonical values. This may be
    unlikely \citep[see e.g.][]{Renaud2017}, in which case this would not
    change the robustness of our inferred slopes against the different BH
    methods we examine here, though more complex tidal histories are possible
    \citep[e.g.][]{Meng2022}. The inclusion of more detailed modelling of the
    evolving MW potential would help increase the confidence of these IMF
    results, however this is out of scope for this work.

\subsection{Black Hole Populations}
\label{sub:black_hole_populations}

    \begin{figure*}
        \centering
        \includegraphics[width=\linewidth]{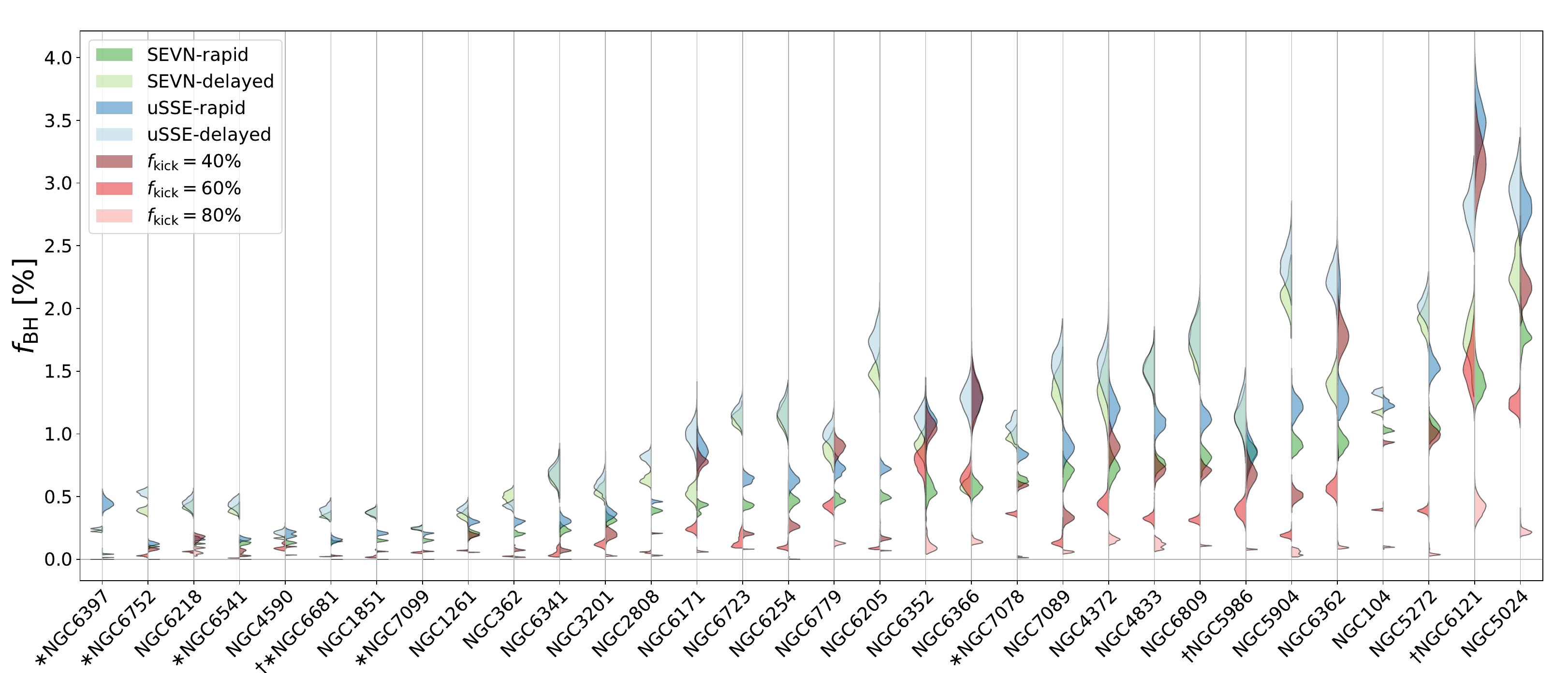}
        \caption{
            Violin plots of the posterior probability distributions of the
            mass fraction in BHs, under each BH prescription explored,
            for all clusters in our sample except \NGC{5139} (\omegacen),
            which has a high inferred \fbh of \(\gtrsim7\%\) under each
            prescription, and is excluded in order to better highlight the
            distributions of the other clusters.
            The delayed SN prescriptions are shown to the left, and the rapid
            SN prescriptions to the right of the vertical line for each cluster.
            Clusters are sorted by the \fbh values of the \SEVNrapid fits.
            All clusters classified as core-collapsed in \citet{Trager1995}
            are denoted by an asterisk. All clusters with
            \(\RGeff<\SI{1.5}{\kilo\pc}\) are denoted by a dagger.
        }
        \label{fig:BH_violins}
    \end{figure*}

    \Cref{fig:BH_violins} shows the inferred posterior probability distributions
    of the present-day mass fractions in BHs (\fbh) from our fits to all
    clusters in our sample, for each BH prescription that we examine.
    While there is a spread in the recovered \fbh values of each cluster
    between the different prescriptions, as with the other present-day
    quantities, for most prescriptions we infer relatively similar values to
    one another.
    This is despite the fact that the various prescriptions begin with
    different initial \fbhi values (see \Cref{sec:bh_methods}).
    In all but the clusters with the highest \fbh values, the median \fbh
    values differ by \(\lesssim 0.5\%\) across prescriptions.
    In many clusters, the \fbh distributions are consistent with one another
    within statistical uncertainties.
    In general, with the delayed SNe prescriptions, we tend to infer
    slightly more BHs than with the rapid SNe prescriptions, and with the
    \uSSE prescription more than with \SEVN, though
    with some notable deviations from these trends in individual clusters.

    As was discussed in \citet{Dickson2024} and \paperI, the uncertainties we
    report here are purely statistical, and likely underestimate the
    true uncertainties.
    The discrepancies seen between the different BH methods in
    \Cref{fig:BH_violins} are likely all smaller than the expected
    true underlying uncertainties.
    The models under each prescription provide good fits to the observations,
    and the available kinematic data near the cluster centres (where
    the strongest signatures of the BHs are expected) is likely not deep or
    precise enough to actually place stronger constraints on \fbh than the
    spread seen across these prescriptions.

    The one set of models that stands out from the rest in many clusters
    is the \fkickeighty models. In all clusters, these fits infer very small
    (\(<0.5\%\)) BH mass fractions. This is directly caused by the very
    small initial BH mass fractions (\(\fbhi<2\%\)), which
    leave little room for larger present-day \fbh values.



\section{Discussion}\label{sec:discussion}

\subsection{Implications for BH physics}
\label{sub:BH_implications}

    We have shown in \Cref{sub:initial_conditions} that a clear relationship
    exists between the assumed BH prescriptions (and thus \fbhi) and the
    inferred initial cluster densities.
    This degeneracy may make it difficult to place direct and stringent
    constraints on the initial densities of clusters based on their present-day
    conditions, as significant uncertainties remain in the choice of BH
    prescriptions employed.
    However, if information about the initial conditions of clusters can be
    provided through other, independent means, this relation could have useful
    implications for the BH physics discussed here.

    In recent years, observations of strongly lensed galaxies with JWST have
    begun to uncover populations of high-redshift candidate proto-GCs, which can
    provide more direct constraints on the early conditions of massive star
    clusters \citep[e.g.][]{Vanzella2023,Claeyssens2023,Adamo2024,
    Claeyssens2026}.
    While substantial uncertainties still exist in the recovered physical
    parameters of these young clusters, which limits the direct comparisons
    we can make with the MW GCs we see today, these observations still allow
    us to place some limits on the expected initial cluster sizes.
    In \paperI, for example, we demonstrated that the inferred densities of most
    clusters (under the \uSSErapid prescriptions), while high, were generally
    consistent with the proto-GCs discovered by \citet{Adamo2024} and
    \citet{Vanzella2023}.
    Here, however, we have found that the initial densities of the \SEVN models
    (using the \PARSEC stellar evolution tracks) are nearly an order of magnitude
    high than those of models using \uSSE, which places them in tension with
    these high-redshift observations.
    In the future, as the catalogues of young proto-GCs grows, the constraints
    they can provide on early cluster densities could be used to place more
    meaningful constraints on the uncertain BH processes which we explore here.

    The appearance of certain features in the distributions of upcoming
    observations of GWs,
    such as a peak in the primary masses within the PISN mass gap,
    may also help place constraints on the initial densities
    of clusters \citep{Ginat2026}, and thus on the underlying BH physics.

    In the other extreme, our models with significant natal kicks (i.e.
    \fkickeighty) result in clusters with very low initial densities,
    somewhat lower present-day densities than seen with other prescriptions,
    and with nearly zero remaining BHs in the
    majority of clusters. Kicks this strong are thus possibly in tension with,
    for example, recent observational discoveries of stellar-mass BHs in GCs
    \citep[e.g.][]{Giesers2018,Giesers2019,Whitaker2026}.

\subsection{Implications for BBH mergers}
\label{sub:BBH_mergers}

    The density and initial BH mass function of star clusters play an important
    role in determining the rate and properties of dynamically-formed BBH
    mergers within them.
    In \paperI we noted that the inferred densities of our sample of
    clusters (\uSSErapid) were consistent with the merger rates and redshift
    dependence seen in recent LIGO-Virgo-KAGRA (LVK) catalogues
    \citep[GWTC-5.0;][]{LIGO2026}, according to the population models of
    \citet{Antonini2020b}. Here we extend this analysis by directly comparing
    the population of mergers occurring in our clusters under different BH
    prescriptions.

    To simulate the BBH mergers we use the BBH dynamics model \BHBdynamics.
    These models comprise the second part of the full
    \textsc{clusterBHBdynamics} (\cBHBd) models, alongside
    \clusterBH, and were first described in \citet{Antonini2020a,Antonini2020b}
    and recently updated in \cbhpaper and \citet{Chattopadhyay2026}.
    \BHBdynamics works alongside \clusterBH by sampling a discrete population
    of BHs from the BH mass function, forming a BBH and considering the few-body
    dynamics of the binary-single and binary-binary interactions it experiences
    over its lifetime, until the BBH merges or is ejected from the cluster,
    and a new BBH is formed.
    Here we use the BBH remnant models of \citet{Varma2019} and
    \citet{Islam2023}.
    For each cluster in our sample, under both the rapid and delayed \uSSE
    and \SEVN prescriptions, we generate the mergers using \BHBdynamics based
    on the initial conditions and \clusterBH evolution we infer.

\begin{figure}
        \centering
        \includegraphics[width=\linewidth]{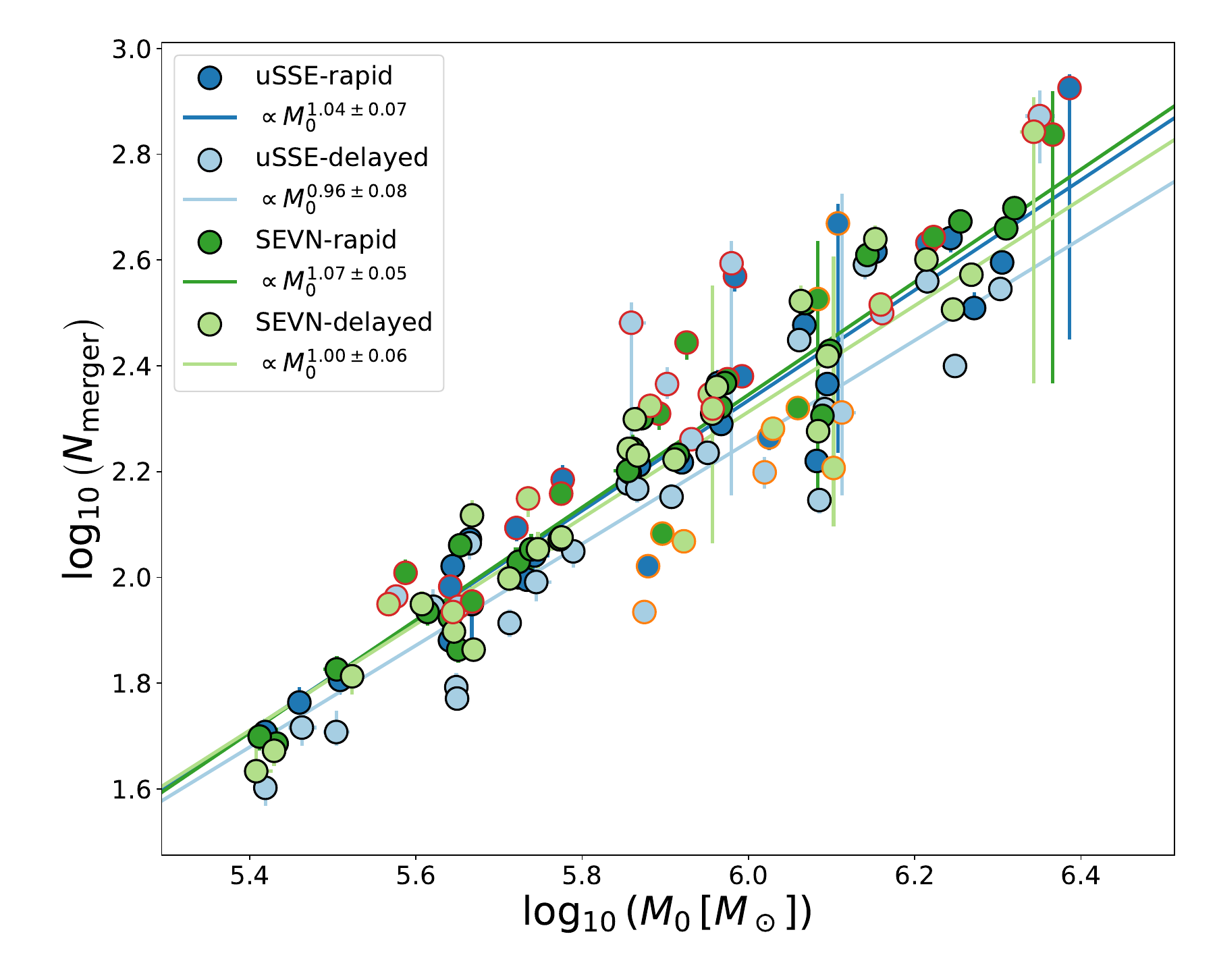}
        \caption{
            Relationship between the initial cluster mass \Mi and the
            total number of dynamical BBH mergers across all clusters, under
            each \uSSE and \SEVN BH prescription explored.
            For each prescription, a best-fitting power-law is also shown.
            All core-collapsed clusters are outlined in red.
            All clusters with \(\RGeff<\SI{1.5}{\kilo\pc}\) are outlined in
            orange.
        }
        \label{fig:M0_Nmergers}
    \end{figure}

    In \Cref{fig:M0_Nmergers} we show the number of mergers (including
    higher generation mergers) produced by each cluster, as a function of
    the initial cluster mass.
    A linear relationship between the number of mergers and initial
    cluster mass is clear across all BH prescriptions.
    This relationship was reported in \cbhpaper (see their Section 4), and our
    results confirm that it is robust against the exact choice of stellar
    evolution models and BH prescriptions.
    Notably, despite the \SEVN models beginning with higher initial densities
    and BH mass fractions, the total number of mergers
    occurring within a given cluster does not differ significantly between the
    \SEVN and \uSSE fits.
    For clusters with the same mass and density, the higher \(m_1\) in \SEVN
    models is expected to lead to a lower merger rate, because the BBH merger
    rate scales as \(\propto m_1^{-1}\) for a given \(q\)
    \citep[][see their Equation 29]{RandoForastier2025}.
    However, the \SEVN models require significantly higher initial densities,
    which increases the BBH merger rate and approximately compensates for this
    effect of \(m_1\).

\begin{figure}
        \centering
        \includegraphics[width=\linewidth]{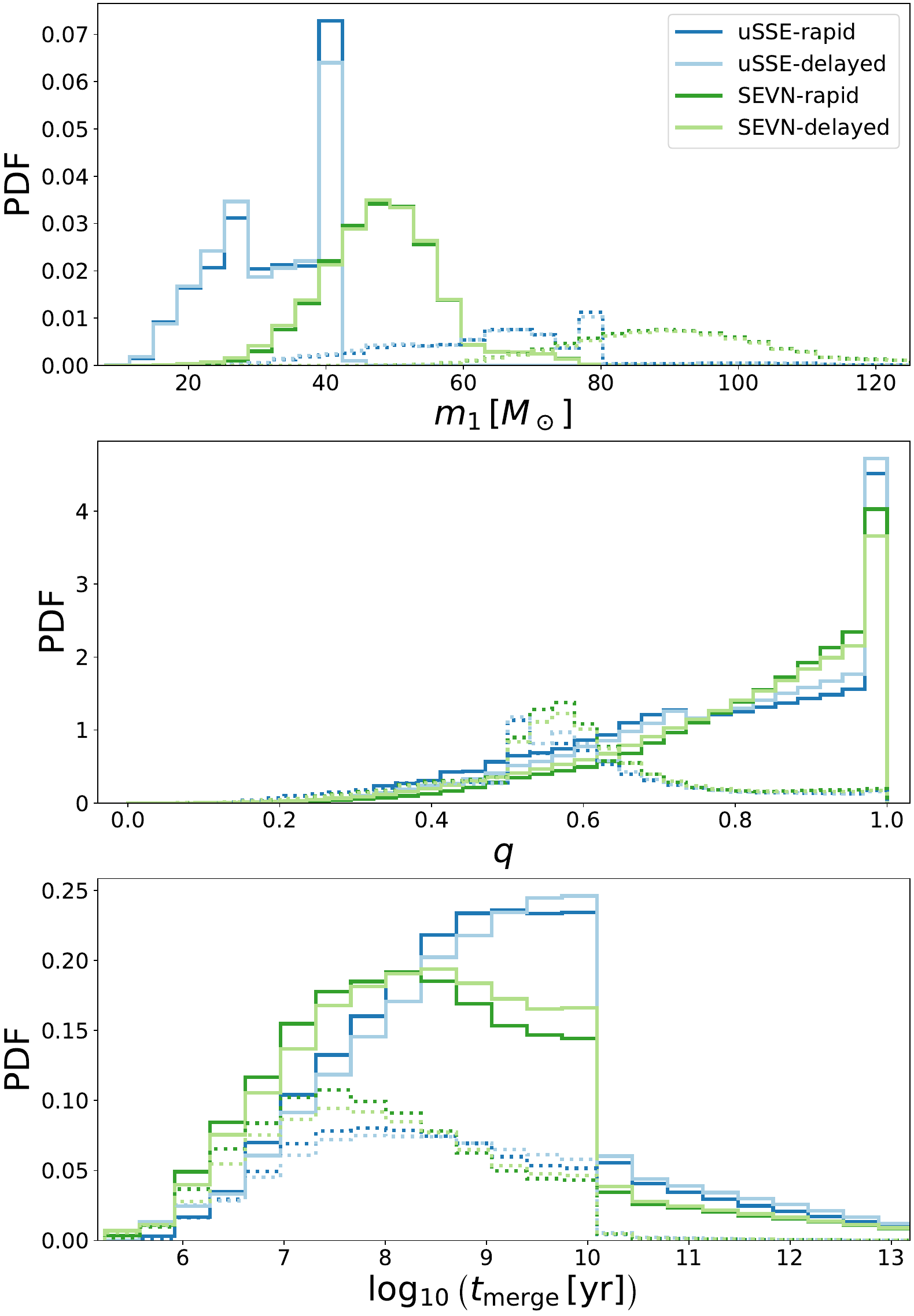}
        \caption{
            The distributions of primary masses \(m_1\) (top panel), mass
            ratios \(q=m_2/m_1\) (middle panel) and merger timescales
            \(t_{\mathrm{merge}}\) (bottom panel)
            of the complete sample of dynamical BBH mergers across all clusters,
            under each \uSSE and \SEVN BH prescription explored.
            All first-generation mergers are shown with solid lines, and all
            higher-generation mergers with dashed lines.
        }
        \label{fig:merger_properties}
    \end{figure}

    While \fbhi is higher in the \SEVN models, this is driven by the higher
    average masses of the BHs formed, and the actual \textit{numbers} of BHs
    initially created and retained after natal kicks are generally similar
    across BH prescriptions (within \(\Delta\Nbh<50\) in almost all clusters).
    This higher average BH mass is also clear in the component masses of these
    mergers, as seen in \Cref{fig:merger_properties}, which shows
    select properties (primary masses, mass ratios and merger times) of the
    total mergers across the sample of clusters.
    This figure also shows that the mass ratios \(q=m_2/m_1\) of the first
    generation mergers are slightly higher on average under \SEVN, and the
    mergers occur at earlier times.
    The \uSSE models have a distribution of primary masses peaking at lower
    values than the \SEVN models (peaking near 30 and \SI{50}{\Msun}
    respectively in the first generation of mergers), reflecting the
    differences in their BH mass functions.
    The \uSSE distribution also shows a spike of mergers at
    \(m_1\sim\SI{40}{\Msun}\), which is caused by
    the pair-instability SN prescriptions of \citet{Belczynski2016} capping
    BH masses from progenitors \(>\SI{100}{\Msun}\) at
    \(\sim\SI{40}{\Msun}\).

    It is important to note that we are examining here a limited sample of
    MW clusters, biased towards nearby and massive GCs, and we are not
    conducting a full population synthesis study.
    The merger properties discussed here reflect only those that we expect
    to occur in these specific MW clusters, and do
    not account for, e.g., a population of lower-mass clusters which may
    have dissolved by the present day.
    If we assume the relationship in
    \Cref{fig:M0_Nmergers} holds to lower masses, and that the underlying
    initial cluster mass function scales as \(\propto M_0^{-2}\)
    \citep{PortegiesZwart2010}, then the approximately constant merger
    efficiency (\(N_{\mathrm{merger}}/M_0\))
    implies that low- and high-mass clusters contribute an equal amount
    to the overall merger rate (as also noted by \cbhpaper).
    Nonetheless, this illustrates how different assumptions
    underlying the BH formation physics, while still leading to nearly identical
    present-day cluster conditions and number of mergers, will display
    different BBH merger properties.

    The impacts of these kinds of stellar evolution and BH physics assumptions
    on the merger rates and GW parameter correlations will be explored in more
    detail in a forthcoming work (F. Fronimos Pouliasis et al., in prep.).

\subsection{Implications for IMBH Formation}
\label{sub:IMBH_formation}

    In \paperI we examined the potential formation of IMBHs, through both
    the runaway stellar collisions and hierarchical BH mergers pathways, by
    applying various relationships found in the literature which predicted IMBH
    masses in GCs formed through these mechanisms
    \citep[e.g.][]{Antonini2019,Vergara2026,Rantala2026,Bocchi2026}.
    It was found that, despite the relatively high initial densities inferred,
    very few, if any, of our clusters could be expected to grow a very
    massive IMBH (\(\gtrsim10^4\,\Msun\)).

    These conclusions still hold for the results presented here for all of the
    \uSSE fits, as well as the \fkick grid which all have lower initial
    densities.
    The fits under \SEVN do have notably higher initial densities, as shown
    in \Cref{sub:initial_conditions}; however these still do not lead to
    significant predicted IMBH masses in most clusters.
    All clusters, except for the same few outliers noted in \paperI
    which should be regarded with some caution, still fall below
    the mass-dependent critical density for forming \(>10^4\,\Msun\)
    IMBHs via runaway collisions presented by \citet{Vergara2026}.
    Applying the IMBH mass fitting formulae from \citet{Rantala2026}
    leads to slightly higher IMBH masses under \SEVN through this mechanism,
    but the maximum predicted values (\(\lesssim \SI{5000}{\Msun}\)) are still
    well below \(10^4\,\Msun\).

\begin{figure}
        \centering
        \includegraphics[width=\linewidth]{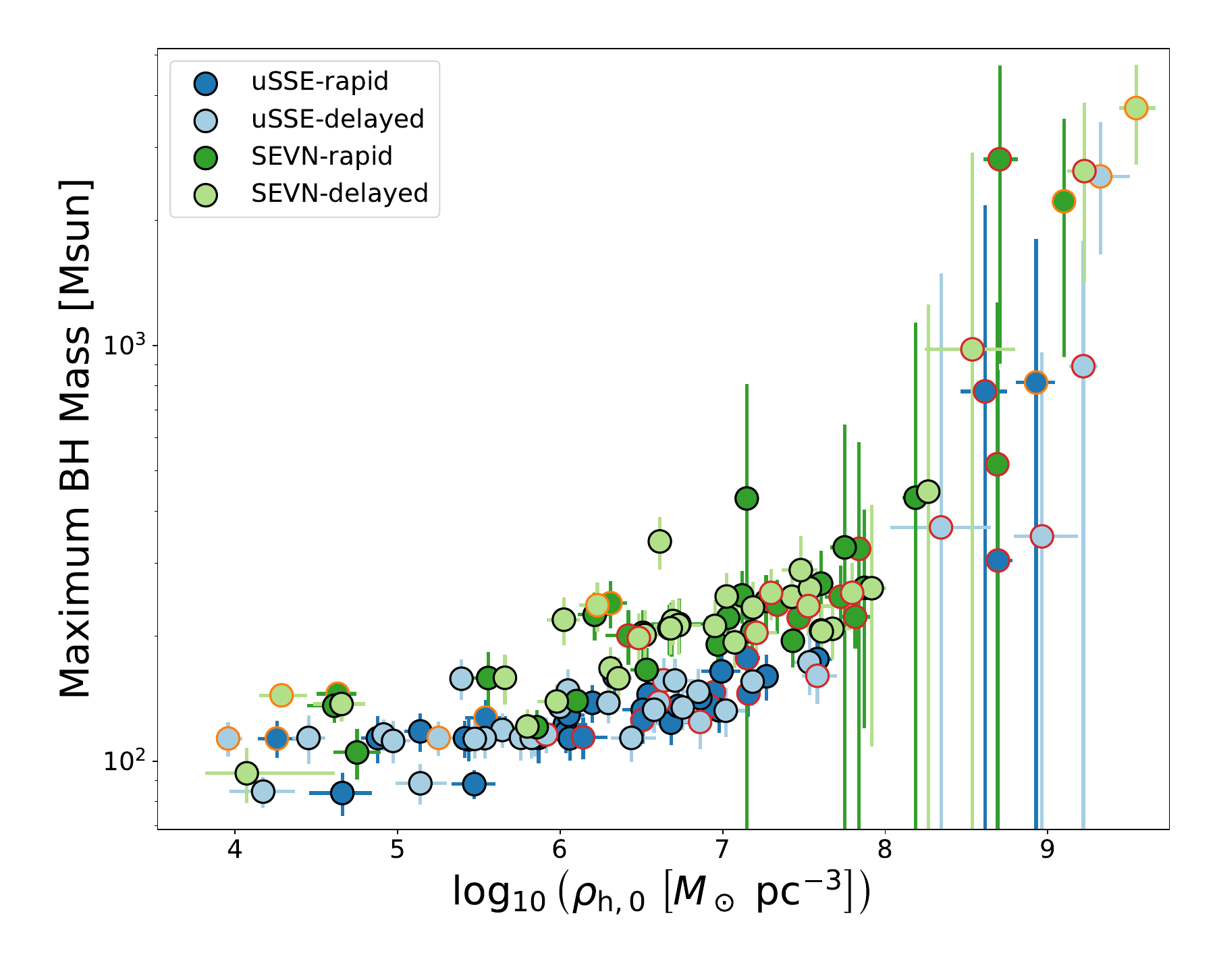}
        \caption{
            The maximum BH mass which is formed through dynamical BBH mergers
            and retained in the cluster after GW recoil kicks, as a function
            of the initial cluster half-mass density, across all
            clusters, under each \uSSE and \SEVN BH prescription explored.
            All core-collapsed clusters are outlined in red.
            All clusters with \(\RGeff<\SI{1.5}{\kilo\pc}\) are outlined in
            orange.
        }
        \label{fig:merger_IMBHs}
    \end{figure}

    For the hierarchical BH mergers pathway, rather than relying on the
    relations presented by \citet{Antonini2019} again, we can examine the BH
    masses which are grown through mergers as computed by the updated version of
    \BHBdynamics (\Cref{sub:BBH_mergers}).
    The maximum BH masses which are grown through BBH mergers and retained in
    the cluster are shown in \Cref{fig:merger_IMBHs} for all clusters in our
    sample, as a function of the inferred initial half-mass densities.
    As might be expected, we find a slight correlation between the most massive
    merger products within a cluster and the density,
    as higher escape velocities allow the mergers to be retained after their GW
    recoil kicks.
    However, under both \SEVN and \uSSE, we find that even the densest clusters
    typically only grow and retain BHs of masses \(\lesssim \SI{500}{\Msun}\).
    The \SEVN models result in slightly higher final BH masses, due to the
    higher average BH masses created under the \SEVN BH IFMR and the higher
    densities we infer.
    This lack of BH growth, despite the generally high densities, occurs
    because the initial escape velocities of most GCs are
    \(\lesssim \SI{300}{\km\per\second}\), that is, below the critical value
    required to retain the first few BBH mergers after their GW kicks
    \citep{Antonini2019}.
    It is not until reaching densities of nearly
    \(10^{9}\, \unit{\Msun \ \pc^{-3}}\) that the largest BH masses approach
    \(10^{3}\, \unit{\Msun}\), and, as discussed above and in \paperI, all of
    these clusters should be regarded with caution.
    This pathway will be explored in more detail, using \cBHBd, in a
    forthcoming work (Marín Pina et al., in prep.).

    In summary, while we may expect stellar collisions and hierarchical
    BH mergers to be a potential avenue for the formation of lower-mass
    IMBHs in some clusters, it is unlikely that this would lead to very
    massive (\(\gtrsim 10^4\,\Msun\)) IMBHs, even in our densest sets of models.

    Note, however, that we have considered here only the results of both
    pathways independently, and in reality some combination of stellar
    collisions providing a larger seed for further, long-term BH growth could
    potentially lead to larger IMBHs forming.



\section{Conclusions}\label{sec:conclusions}

    In this paper, we have extended the work of \paperI, fitting new coupled
    fast evolutionary and multimass equilibrium models to a sample of 35 MW
    GCs, by exploring the impacts of a suite of prescriptions for
    BH IFMRs and natal kicks on the inferred initial conditions.
    We summarize our main conclusions here:

    \begin{enumerate}

        \item
        Under the various BH prescriptions examined, the best-fitting models
        for each cluster result in very similar present-day conditions, despite
        beginning with notably different BH mass fractions. This is achieved
        by variations in the inferred initial half-mass radii/densities, while
        the inferred initial masses remain very similar across methods.

        \item
        The inferred distributions of initial half-mass densities
        (\(\log_{10}(\rhohi)\)) across our
        sample of clusters follow an approximately linear relationship with
        the initial BH mass fractions (\fbhi) under a given set of BH
        prescriptions.

        \item
        The inferred initial conditions are largely insensitive to the
        choice of rapid or delayed SN mechanism, as the total change in initial
        BH mass fractions between the two is small.
        The \PARSEC stellar evolution tracks (implemented in \SEVN) result in a
        distribution of initial densities nearly an order of magnitude higher
        than under the prescriptions of \uSSE.

        \item
        We recover nearly identical, bottom-light stellar IMFs across the
        different BH prescriptions, highlighting the robustness of this
        result to the chosen underlying physics.

        \item
        As in \paperI, we infer relatively small populations of BHs at the
        present day (typically less than \(\fbh=1.5\) per cent) under all
        BH prescriptions explored, though with a notable spread between the
        methods within individual clusters, reflecting the level of precision
        that can be expected of our methods.

        \item
        We find that the total number of BBH mergers expected within a given
        cluster is not sensitive to the chosen BH prescription, though the
        properties of the mergers themselves are, with the \SEVN models
        predicting earlier mergers, with higher primary masses.

        \item
        Despite the even higher initial densities inferred under the \SEVN
        models, we still do not predict the formation of very massive
        IMBHs \(M_{\mathrm{IMBH}}\gtrsim10^4\,\Msun\), through either runaway
        stellar collisions or hierarchical BH mergers, in most clusters.

    \end{enumerate}

\section*{Acknowledgements}

    ND is grateful for the support of the Durland Scholarship in Graduate
    Research and the MITACS Globalink Research Internship program, and for the
    hospitality of ICCUB during the completion of this work.
    VHB acknowledges the support of the Natural Sciences and Engineering
    Research Council of Canada (NSERC) through grant RGPIN-2020-05990.
    FFP acknowledges the “la Caixa” Foundation (ID100010434) for financial
    support in the form of a Doctoral INPhINIT fellowship (fellowship code
    LCF/BQ/DI23/11990067). MG acknowledges financial support
    from the grants  PID2024-155720NB-I00, CEX2024-001451-M funded by
    MCIN/AEI/10.13039/501100011033 (State Agency for Research of the
    Spanish Ministry of Science and Innovation). DMP acknowledges support
    from the Deutsche Forschungsgemeinschaft (DFG, German Research Foundation)
    through project number 546850815 (acronym: DoBlack) and under Germany's
    Excellence Strategy EXC 2181/1 - 390900948 (the Heidelberg STRUCTURES
    Excellence Cluster).

    This research was enabled in part by support provided by ACENET
    (\url{www.ace-net.ca}) and the Digital Research Alliance of Canada
    (\url{https://alliancecan.ca}).

    \software{
        \texttt{astropy} \citep{Astropy2018},
        \texttt{dynesty} \citep{Speagle2020,Koposov2025},
        \texttt{h5py} \citep{Collette2022},
        \texttt{cmctoolkit} \citep{cmctoolkit},
        \texttt{matplotlib} \citep{Hunter2007},
        \texttt{numpy} \citep{Harris2020},
        \texttt{scipy} \citep{Virtanen2020}
    }


    \bibliographystyle{aasjournal}
    \bibliography{biblio}

@ARTICLE{Dickson2023,
       author = {{Dickson}, N. and {H{\'e}nault-Brunet}, V. and {Baumgardt}, H. and {Gieles}, M. and {Smith}, P.~J.},
        title = "{Multimass modelling of Milky Way globular clusters - I. Implications on their stellar initial mass function above 1 M$_{{\ensuremath{\odot}}}$}",
      journal = {\mnras},
         year = 2023,
        month = jul,
       volume = {522},
       number = {4},
        pages = {5320-5339},
          doi = {10.1093/mnras/stad1254},
archivePrefix = {arXiv},
       eprint = {2303.01637},
 primaryClass = {astro-ph.GA},
       adsurl = {https://ui.adsabs.harvard.edu/abs/2023MNRAS.522.5320D}
}

@ARTICLE{Fronimos2026,
       author = {{Fronimos Pouliasis}, Fotios and {Dickson}, Nolan and {Mar{\'\i}n Pina}, Daniel and {Gieles}, Mark and {H{\'e}nault-Brunet}, Vincent and {Antonini}, Fabio},
        title = "{cBHBd: A fast code for the evolution of tidally limited star clusters and their binary black hole mergers}",
      journal = {arXiv e-prints},
         year = 2026,
        month = may,
          eid = {arXiv:2605.28088},
        pages = {arXiv:2605.28088},
archivePrefix = {arXiv},
       eprint = {2605.28088},
 primaryClass = {astro-ph.GA},
       adsurl = {https://ui.adsabs.harvard.edu/abs/2026arXiv260528088F}
}

@ARTICLE{Dickson2026a,
       author = {{Dickson}, Nolan and {H{\'e}nault-Brunet}, Vincent and {Fronimos Pouliasis}, Fotios and {Gieles}, Mark and {Smith}, Peter J.},
        title = "{Fast Dynamical Modelling of Milky Way Globular Clusters -- I. Implications for Initial Cluster Densities}",
      journal = {arXiv e-prints},
         year = 2026,
        month = jul,
          eid = {arXiv:2607.07010},
        pages = {arXiv:2607.07010},
          doi = {10.48550/arXiv.2607.07010},
archivePrefix = {arXiv},
       eprint = {2607.07010},
 primaryClass = {astro-ph.GA},
       adsurl = {https://ui.adsabs.harvard.edu/abs/2026arXiv260707010D}
}

@ARTICLE{Dickson2024,
       author = {{Dickson}, N. and {Smith}, P.~J. and {H{\'e}nault-Brunet}, V. and {Gieles}, M. and {Baumgardt}, H.},
        title = "{Multimass modelling of milky way globular clusters - II. Present-day black hole populations}",
      journal = {\mnras},
         year = 2024,
        month = mar,
       volume = {529},
       number = {1},
        pages = {331-347},
          doi = {10.1093/mnras/stae470},
archivePrefix = {arXiv},
       eprint = {2308.13037},
 primaryClass = {astro-ph.GA},
       adsurl = {https://ui.adsabs.harvard.edu/abs/2024MNRAS.529..331D}
}

@article{Alexander2012,
  title = {A Prescription and Fast Code for the Long-Term Evolution of Star Clusters: {{A}} Code for the Evolution of Star Clusters},
  shorttitle = {A Prescription and Fast Code for the Long-Term Evolution of Star Clusters},
  author = {Alexander, Poul E. R. and Gieles, Mark},
  year = 2012,
  month = jun,
  journal = {Monthly Notices of the Royal Astronomical Society},
  volume = {422},
  number = {4},
  pages = {3415--3432},
  doi = {10.1111/j.1365-2966.2012.20867.x}
}

@article{Gieles2014,
  title = {A Prescription and Fast Code for the Long-Term Evolution of Star Clusters -- {{II}}. {{Unbalanced}} and Core Evolution},
  author = {Gieles, Mark and Alexander, Poul E. R. and Lamers, Henny J. G. L. M. and Baumgardt, Holger},
  year = 2014,
  month = jan,
  journal = {Monthly Notices of the Royal Astronomical Society},
  volume = {437},
  number = {1},
  pages = {916--929},
  doi = {10.1093/mnras/stt1980}
}

@article{Alexander2014,
  title = {A Prescription and Fast Code for the Long-Term Evolution of Star Clusters - {{III}}. {{Unequal}} Masses and Stellar Evolution},
  author = {Alexander, Poul E. R. and Gieles, Mark and Lamers, Henny J. G. L. M. and Baumgardt, Holger},
  year = 2014,
  month = aug,
  journal = {Monthly Notices of the Royal Astronomical Society},
  volume = {442},
  pages = {1265--1285},
  publisher = {OUP},
  doi = {10.1093/mnras/stu899}
}

@ARTICLE{Chattopadhyay2026,
       author = {{Chattopadhyay}, Debatri and {Mar{\'\i}n Pina}, Daniel and {Gieles}, Mark and {Antonini}, Fabio and {Fronimos Pouliasis}, Fotios},
        title = "{Repopulating the pair-instability mass gap without sustained growth to massive IMBHs: the case of 47\textbackslash,Tuc}",
      journal = {arXiv e-prints},
         year = 2026,
        month = apr,
          eid = {arXiv:2604.09773},
        pages = {arXiv:2604.09773},
          doi = {10.48550/arXiv.2604.09773},
archivePrefix = {arXiv},
       eprint = {2604.09773},
 primaryClass = {astro-ph.HE},
       adsurl = {https://ui.adsabs.harvard.edu/abs/2026arXiv260409773C}
}

@ARTICLE{Islam2023,
       author = {{Islam}, Tousif and {Field}, Scott E. and {Khanna}, Gaurav},
        title = "{Remnant black hole properties from numerical-relativity-informed perturbation theory and implications for waveform modeling}",
      journal = {\prd},
         year = 2023,
        month = sep,
       volume = {108},
       number = {6},
          eid = {064048},
        pages = {064048},
          doi = {10.1103/PhysRevD.108.064048},
archivePrefix = {arXiv},
       eprint = {2301.07215},
 primaryClass = {gr-qc},
       adsurl = {https://ui.adsabs.harvard.edu/abs/2023PhRvD.108f4048I}
}

@ARTICLE{Kritos2024,
       author = {{Kritos}, Konstantinos and {Strokov}, Vladimir and {Baibhav}, Vishal and {Berti}, Emanuele},
        title = "{Dynamical formation of black hole binaries in dense star clusters: Rapid cluster evolution code}",
      journal = {\prd},
         year = 2024,
        month = aug,
       volume = {110},
       number = {4},
          eid = {043023},
        pages = {043023},
          doi = {10.1103/PhysRevD.110.043023},
archivePrefix = {arXiv},
       eprint = {2210.10055},
 primaryClass = {astro-ph.HE},
       adsurl = {https://ui.adsabs.harvard.edu/abs/2024PhRvD.110d3023K}
}

@ARTICLE{Mapelli2021,
       author = {{Mapelli}, Michela and {Dall'Amico}, Marco and {Bouffanais}, Yann and {Giacobbo}, Nicola and {Arca Sedda}, Manuel and {Artale}, M. Celeste and {Ballone}, Alessandro and {Di Carlo}, Ugo N. and {Iorio}, Giuliano and {Santoliquido}, Filippo and {Torniamenti}, Stefano},
        title = "{Hierarchical black hole mergers in young, globular and nuclear star clusters: the effect of metallicity, spin and cluster properties}",
      journal = {\mnras},
         year = 2021,
        month = jul,
       volume = {505},
       number = {1},
        pages = {339-358},
          doi = {10.1093/mnras/stab1334},
archivePrefix = {arXiv},
       eprint = {2103.05016},
 primaryClass = {astro-ph.HE},
       adsurl = {https://ui.adsabs.harvard.edu/abs/2021MNRAS.505..339M}
}

@ARTICLE{Gieles2015,
       author = {{Gieles}, Mark and {Zocchi}, Alice},
        title = "{A family of lowered isothermal models}",
      journal = {\mnras},
         year = 2015,
        month = nov,
       volume = {454},
       number = {1},
        pages = {576-592},
          doi = {10.1093/mnras/stv1848},
archivePrefix = {arXiv},
       eprint = {1508.02120},
 primaryClass = {astro-ph.IM},
       adsurl = {https://ui.adsabs.harvard.edu/abs/2015MNRAS.454..576G}
}

@article{Balbinot2018,
  title = {The Devil Is in the Tails: The Role of Globular Cluster Mass Evolution on Stream Properties},
  shorttitle = {The Devil Is in the Tails},
  author = {Balbinot, Eduardo and Gieles, Mark},
  year = 2018,
  month = feb,
  journal = {Monthly Notices of the Royal Astronomical Society},
  volume = {474},
  pages = {2479--2492},
  publisher = {OUP},
  doi = {10.1093/mnras/stx2708}
}

@ARTICLE{Henault-Brunet2020,
       author = {{H{\'e}nault-Brunet}, V. and {Gieles}, M. and {Strader}, J. and {Peuten}, M. and {Balbinot}, E. and {Douglas}, K.~E.~K.},
        title = "{On the black hole content and initial mass function of 47 Tuc}",
      journal = {\mnras},
         year = 2020,
        month = jan,
       volume = {491},
       number = {1},
        pages = {113-128},
          doi = {10.1093/mnras/stz2995},
archivePrefix = {arXiv},
       eprint = {1908.08538},
 primaryClass = {astro-ph.GA},
       adsurl = {https://ui.adsabs.harvard.edu/abs/2020MNRAS.491..113H}
}

@article{Gieles2021,
  title = {A Supra-Massive Population of Stellar-Mass Black Holes in the Globular Cluster {{Palomar}} 5},
  author = {Gieles, Mark and Erkal, Denis and Antonini, Fabio and Balbinot, Eduardo and Pe{\~n}arrubia, Jorge},
  year = 2021,
  month = jul,
  journal = {Nature Astronomy},
  volume = {5},
  number = {9},
  eprint = {2102.11348},
  primaryclass = {astro-ph},
  pages = {957--966},
  doi = {10.1038/s41550-021-01392-2},
  archiveprefix = {arXiv}
}

@ARTICLE{Kremer2020,
       author = {{Kremer}, Kyle and {Ye}, Claire S. and {Rui}, Nicholas Z. and {Weatherford}, Newlin C. and {Chatterjee}, Sourav and {Fragione}, Giacomo and {Rodriguez}, Carl L. and {Spera}, Mario and {Rasio}, Frederic A.},
        title = "{Modeling Dense Star Clusters in the Milky Way and Beyond with the CMC Cluster Catalog}",
      journal = {\apjs},
         year = 2020,
        month = apr,
       volume = {247},
       number = {2},
          eid = {48},
        pages = {48},
          doi = {10.3847/1538-4365/ab7919},
archivePrefix = {arXiv},
       eprint = {1911.00018},
 primaryClass = {astro-ph.HE},
       adsurl = {https://ui.adsabs.harvard.edu/abs/2020ApJS..247...48K}
}

@INPROCEEDINGS{Kremer2020b,
       author = {{Kremer}, Kyle and {Ye}, Claire S. and {Chatterjee}, Sourav and {Rodriguez}, Carl L. and {Rasio}, Frederic A.},
        title = "{The Role of ``black hole burning'' in the evolution of dense star clusters}",
       series = "",
    booktitle = {Star Clusters: From the Milky Way to the Early Universe},
         year = 2020,
       editor = {{Bragaglia}, Angela and {Davies}, Melvyn and {Sills}, Alison and {Vesperini}, Enrico},
       volume = {351},
        month = jan,
        pages = {357-366},
          doi = {10.1017/S1743921319007269},
}

@ARTICLE{Rodriguez2022,
       author = {{Rodriguez}, Carl L. and {Weatherford}, Newlin C. and {Coughlin}, Scott C. and {Amaro-Seoane}, Pau and {Breivik}, Katelyn and {Chatterjee}, Sourav and {Fragione}, Giacomo and {K{\i}ro{\u{g}}lu}, Fulya and {Kremer}, Kyle and {Rui}, Nicholas Z. and {Ye}, Claire S. and {Zevin}, Michael and {Rasio}, Frederic A.},
        title = "{Modeling Dense Star Clusters in the Milky Way and beyond with the Cluster Monte Carlo Code}",
      journal = {\apjs},
         year = 2022,
        month = feb,
       volume = {258},
       number = {2},
          eid = {22},
        pages = {22},
          doi = {10.3847/1538-4365/ac2edf},
archivePrefix = {arXiv},
       eprint = {2106.02643},
 primaryClass = {astro-ph.GA},
       adsurl = {https://ui.adsabs.harvard.edu/abs/2022ApJS..258...22R}
}

@inproceedings{Henon1975,
  title = {Two {{Recent Developments Concerning}} the {{Monte Carlo Method}}},
  booktitle = {Dynamics of {{Stellar Systems}}},
  author = {H{\'e}non, M.},
  year = 1975,
  month = jan,
  volume = {69},
  pages = {133}
}

@ARTICLE{Trager1995,
       author = {{Trager}, S.~C. and {King}, Ivan R. and {Djorgovski}, S.},
        title = "{Catalogue of Galactic Globular-Cluster Surface-Brightness Profiles}",
      journal = {\aj},
         year = 1995,
        month = jan,
       volume = {109},
        pages = {218},
          doi = {10.1086/117268},
       adsurl = {https://ui.adsabs.harvard.edu/abs/1995AJ....109..218T}
}

@ARTICLE{Baumgardt2018,
       author = {{Baumgardt}, H. and {Hilker}, M.},
        title = "{A catalogue of masses, structural parameters, and velocity dispersion profiles of 112 Milky Way globular clusters}",
      journal = {\mnras},
         year = 2018,
        month = aug,
       volume = {478},
       number = {2},
        pages = {1520-1557},
          doi = {10.1093/mnras/sty1057},
archivePrefix = {arXiv},
       eprint = {1804.08359},
 primaryClass = {astro-ph.GA},
       adsurl = {https://ui.adsabs.harvard.edu/abs/2018MNRAS.478.1520B}
}

@ARTICLE{Baumgardt2023,
       author = {{Baumgardt}, H. and {H{\'e}nault-Brunet}, V. and {Dickson}, N. and {Sollima}, A.},
        title = "{Evidence for a bottom-light initial mass function in massive star clusters}",
      journal = {\mnras},
         year = 2023,
        month = may,
       volume = {521},
       number = {3},
        pages = {3991-4008},
          doi = {10.1093/mnras/stad631},
archivePrefix = {arXiv},
       eprint = {2303.01636},
 primaryClass = {astro-ph.GA},
       adsurl = {https://ui.adsabs.harvard.edu/abs/2023MNRAS.521.3991B}
}

@article{King1966,
  title = {The Structure of Star Clusters. {{III}}. {{Some}} Simple Dynamical Models},
  author = {King, Ivan R.},
  year = 1966,
  month = feb,
  journal = {The Astronomical Journal},
  volume = {71},
  pages = {64},
  publisher = {IOP},
  doi = {10.1086/109857}
}

@ARTICLE{Kroupa2001,
       author = {{Kroupa}, Pavel},
        title = "{On the variation of the initial mass function}",
      journal = {\mnras},
         year = 2001,
        month = apr,
       volume = {322},
       number = {2},
        pages = {231-246},
          doi = {10.1046/j.1365-8711.2001.04022.x},
}

@ARTICLE{Banerjee2020,
       author = {{Banerjee}, S. and {Belczynski}, K. and {Fryer}, C.~L. and {Berczik}, P. and {Hurley}, J.~R. and {Spurzem}, R. and {Wang}, L.},
        title = "{BSE versus StarTrack: Implementations of new wind, remnant-formation, and natal-kick schemes in NBODY7 and their astrophysical consequences}",
      journal = {\aap},
         year = 2020,
        month = jul,
       volume = {639},
          eid = {A41},
        pages = {A41},
          doi = {10.1051/0004-6361/201935332},
archivePrefix = {arXiv},
       eprint = {1902.07718},
 primaryClass = {astro-ph.SR},
       adsurl = {https://ui.adsabs.harvard.edu/abs/2020A&A...639A..41B}
}

@ARTICLE{Aarseth2012,
       author = {{Aarseth}, Sverre J.},
        title = "{Mergers and ejections of black holes in globular clusters}",
      journal = {\mnras},
         year = 2012,
        month = may,
       volume = {422},
       number = {1},
        pages = {841-848},
          doi = {10.1111/j.1365-2966.2012.20666.x},
archivePrefix = {arXiv},
       eprint = {1202.4688},
 primaryClass = {astro-ph.SR},
       adsurl = {https://ui.adsabs.harvard.edu/abs/2012MNRAS.422..841A}
}

@ARTICLE{Hurley2000,
       author = {{Hurley}, Jarrod R. and {Pols}, Onno R. and {Tout}, Christopher A.},
        title = "{Comprehensive analytic formulae for stellar evolution as a function of mass and metallicity}",
      journal = {\mnras},
         year = 2000,
        month = jul,
       volume = {315},
       number = {3},
        pages = {543-569},
          doi = {10.1046/j.1365-8711.2000.03426.x},
archivePrefix = {arXiv},
       eprint = {astro-ph/0001295},
 primaryClass = {astro-ph},
       adsurl = {https://ui.adsabs.harvard.edu/abs/2000MNRAS.315..543H}
}

@ARTICLE{Hurley2002,
       author = {{Hurley}, Jarrod R. and {Tout}, Christopher A. and {Pols}, Onno R.},
        title = "{Evolution of binary stars and the effect of tides on binary populations}",
      journal = {\mnras},
         year = 2002,
        month = feb,
       volume = {329},
       number = {4},
        pages = {897-928},
          doi = {10.1046/j.1365-8711.2002.05038.x},
archivePrefix = {arXiv},
       eprint = {astro-ph/0201220},
 primaryClass = {astro-ph},
       adsurl = {https://ui.adsabs.harvard.edu/abs/2002MNRAS.329..897H}
}

@ARTICLE{Belczynski2010,
       author = {{Belczynski}, Krzysztof and {Bulik}, Tomasz and {Fryer}, Chris L. and {Ruiter}, Ashley and {Valsecchi}, Francesca and {Vink}, Jorick S. and {Hurley}, Jarrod R.},
        title = "{On the Maximum Mass of Stellar Black Holes}",
      journal = {\apj},
         year = 2010,
        month = may,
       volume = {714},
       number = {2},
        pages = {1217-1226},
          doi = {10.1088/0004-637X/714/2/1217},
archivePrefix = {arXiv},
       eprint = {0904.2784},
 primaryClass = {astro-ph.SR},
       adsurl = {https://ui.adsabs.harvard.edu/abs/2010ApJ...714.1217B}
}

@ARTICLE{Belczynski2016,
       author = {{Belczynski}, K. and {Heger}, A. and {Gladysz}, W. and {Ruiter}, A.~J. and {Woosley}, S. and {Wiktorowicz}, G. and {Chen}, H.-Y. and {Bulik}, T. and {O'Shaughnessy}, R. and {Holz}, D.~E. and {Fryer}, C.~L. and {Berti}, E.},
        title = "{The effect of pair-instability mass loss on black-hole mergers}",
      journal = {\aap},
         year = 2016,
        month = oct,
       volume = {594},
          eid = {A97},
        pages = {A97},
          doi = {10.1051/0004-6361/201628980},
archivePrefix = {arXiv},
       eprint = {1607.03116},
 primaryClass = {astro-ph.HE},
       adsurl = {https://ui.adsabs.harvard.edu/abs/2016A&A...594A..97B}
}

@article{Fryer2009,
  title = {Neutrinos from {{Fallback}} onto {{Newly Formed Neutron Stars}}},
  author = {Fryer, Chris L.},
  year = 2009,
  month = jul,
  journal = {The Astrophysical Journal},
  volume = {699},
  pages = {409--420},
  publisher = {IOP},
  doi = {10.1088/0004-637X/699/1/409}
}

@ARTICLE{Fryer2012,
       author = {{Fryer}, Chris L. and {Belczynski}, Krzysztof and {Wiktorowicz}, Grzegorz and {Dominik}, Michal and {Kalogera}, Vicky and {Holz}, Daniel E.},
        title = "{Compact Remnant Mass Function: Dependence on the Explosion Mechanism and Metallicity}",
      journal = {\apj},
         year = 2012,
        month = apr,
       volume = {749},
       number = {1},
          eid = {91},
        pages = {91},
          doi = {10.1088/0004-637X/749/1/91},
archivePrefix = {arXiv},
       eprint = {1110.1726},
 primaryClass = {astro-ph.SR},
       adsurl = {https://ui.adsabs.harvard.edu/abs/2012ApJ...749...91F}
}

@ARTICLE{Riley2022,
       author = {{Riley}, Jeff and {Agrawal}, Poojan and {Barrett}, Jim W. and {Boyett}, Kristan N.~K. and {Broekgaarden}, Floor S. and {Chattopadhyay}, Debatri and {Gaebel}, Sebastian M. and {Gittins}, Fabian and {Hirai}, Ryosuke and {Howitt}, George and {Justham}, Stephen and {Khandelwal}, Lokesh and {Kummer}, Floris and {Lau}, Mike Y.~M. and {Mandel}, Ilya and {de Mink}, Selma E. and {Neijssel}, Coenraad and {Riley}, Tim and {van Son}, Lieke and {Stevenson}, Simon and {Vigna-G{\'o}mez}, Alejandro and {Vinciguerra}, Serena and {Wagg}, Tom and {Willcox}, Reinhold and {Team Compas}},
        title = "{Rapid Stellar and Binary Population Synthesis with COMPAS}",
      journal = {\apjs},
         year = 2022,
        month = feb,
       volume = {258},
       number = {2},
          eid = {34},
        pages = {34},
          doi = {10.3847/1538-4365/ac416c},
archivePrefix = {arXiv},
       eprint = {2109.10352},
 primaryClass = {astro-ph.IM},
       adsurl = {https://ui.adsabs.harvard.edu/abs/2022ApJS..258...34R}
}

@ARTICLE{Breivik2020,
       author = {{Breivik}, Katelyn and {Coughlin}, Scott and {Zevin}, Michael and {Rodriguez}, Carl L. and {Kremer}, Kyle and {Ye}, Claire S. and {Andrews}, Jeff J. and {Kurkowski}, Michael and {Digman}, Matthew C. and {Larson}, Shane L. and {Rasio}, Frederic A.},
        title = "{COSMIC Variance in Binary Population Synthesis}",
      journal = {\apj},
         year = 2020,
        month = jul,
       volume = {898},
       number = {1},
          eid = {71},
        pages = {71},
          doi = {10.3847/1538-4357/ab9d85},
archivePrefix = {arXiv},
       eprint = {1911.00903},
 primaryClass = {astro-ph.HE},
       adsurl = {https://ui.adsabs.harvard.edu/abs/2020ApJ...898...71B}
}

@ARTICLE{Giacobbo2018,
       author = {{Giacobbo}, Nicola and {Mapelli}, Michela and {Spera}, Mario},
        title = "{Merging black hole binaries: the effects of progenitor's metallicity, mass-loss rate and Eddington factor}",
      journal = {\mnras},
         year = 2018,
        month = mar,
       volume = {474},
       number = {3},
        pages = {2959-2974},
          doi = {10.1093/mnras/stx2933},
archivePrefix = {arXiv},
       eprint = {1711.03556},
 primaryClass = {astro-ph.SR},
       adsurl = {https://ui.adsabs.harvard.edu/abs/2018MNRAS.474.2959G}
}

@ARTICLE{Bressan2012,
       author = {{Bressan}, Alessandro and {Marigo}, Paola and {Girardi}, L{\'e}o. and {Salasnich}, Bernardo and {Dal Cero}, Claudia and {Rubele}, Stefano and {Nanni}, Ambra},
        title = "{PARSEC: stellar tracks and isochrones with the PAdova and TRieste Stellar Evolution Code}",
      journal = {\mnras},
         year = 2012,
        month = nov,
       volume = {427},
       number = {1},
        pages = {127-145},
          doi = {10.1111/j.1365-2966.2012.21948.x},
archivePrefix = {arXiv},
       eprint = {1208.4498},
 primaryClass = {astro-ph.SR},
       adsurl = {https://ui.adsabs.harvard.edu/abs/2012MNRAS.427..127B}
}

@ARTICLE{Tang2014,
       author = {{Tang}, Jing and {Bressan}, Alessandro and {Rosenfield}, Philip and {Slemer}, Alessandra and {Marigo}, Paola and {Girardi}, L{\'e}o and {Bianchi}, Luciana},
        title = "{New PARSEC evolutionary tracks of massive stars at low metallicity: testing canonical stellar evolution in nearby star-forming dwarf galaxies}",
      journal = {\mnras},
         year = 2014,
        month = dec,
       volume = {445},
       number = {4},
        pages = {4287-4305},
          doi = {10.1093/mnras/stu2029},
archivePrefix = {arXiv},
       eprint = {1410.1745},
 primaryClass = {astro-ph.SR},
       adsurl = {https://ui.adsabs.harvard.edu/abs/2014MNRAS.445.4287T}
}

@ARTICLE{Chen2015,
       author = {{Chen}, Yang and {Bressan}, Alessandro and {Girardi}, L{\'e}o and {Marigo}, Paola and {Kong}, Xu and {Lanza}, Antonio},
        title = "{PARSEC evolutionary tracks of massive stars up to 350 M$_{☉}$ at metallicities 0.0001 {\ensuremath{\leq}} Z {\ensuremath{\leq}} 0.04}",
      journal = {\mnras},
         year = 2015,
        month = sep,
       volume = {452},
       number = {1},
        pages = {1068-1080},
          doi = {10.1093/mnras/stv1281},
archivePrefix = {arXiv},
       eprint = {1506.01681},
 primaryClass = {astro-ph.SR},
       adsurl = {https://ui.adsabs.harvard.edu/abs/2015MNRAS.452.1068C}
}

@ARTICLE{Hobbs2005,
       author = {{Hobbs}, G. and {Lorimer}, D.~R. and {Lyne}, A.~G. and {Kramer}, M.},
        title = "{A statistical study of 233 pulsar proper motions}",
      journal = {\mnras},
         year = 2005,
        month = jul,
       volume = {360},
       number = {3},
        pages = {974-992},
          doi = {10.1111/j.1365-2966.2005.09087.x},
}

@ARTICLE{Pfahl2002,
       author = {{Pfahl}, Eric and {Rappaport}, Saul and {Podsiadlowski}, Philipp},
        title = "{A Comprehensive Study of Neutron Star Retention in Globular Clusters}",
      journal = {\apj},
         year = 2002,
        month = jul,
       volume = {573},
       number = {1},
        pages = {283-305},
          doi = {10.1086/340494},
}

@article{Disberg2025,
  title = {The {{Kick Velocity Distribution}} of {{Isolated Neutron Stars}}},
  author = {Disberg, Paul and Mandel, Ilya},
  year = 2025,
  month = aug,
  journal = {The Astrophysical Journal},
  volume = {989},
  pages = {L8},
  publisher = {IOP},
  doi = {10.3847/2041-8213/adf286},
}

@article{Iorio2023,
  title = {Compact Object Mergers: Exploring Uncertainties from Stellar and Binary Evolution with {{SEVN}}},
  shorttitle = {Compact Object Mergers},
  author = {Iorio, Giuliano and Mapelli, Michela and Costa, Guglielmo and Spera, Mario and Escobar, Gast{\'o}n J. and Sgalletta, Cecilia and Trani, Alessandro A. and Korb, Erika and Santoliquido, Filippo and Dall'Amico, Marco and Gaspari, Nicola and Bressan, Alessandro},
  year = 2023,
  month = sep,
  journal = {Monthly Notices of the Royal Astronomical Society},
  volume = {524},
  pages = {426--470},
  publisher = {OUP},
  doi = {10.1093/mnras/stad1630}
}

@ARTICLE{Spera2019,
       author = {{Spera}, Mario and {Mapelli}, Michela and {Giacobbo}, Nicola and {Trani}, Alessandro A. and {Bressan}, Alessandro and {Costa}, Guglielmo},
        title = "{Merging black hole binaries with the SEVN code}",
      journal = {\mnras},
         year = 2019,
        month = may,
       volume = {485},
       number = {1},
        pages = {889-907},
          doi = {10.1093/mnras/stz359},
archivePrefix = {arXiv},
       eprint = {1809.04605},
 primaryClass = {astro-ph.HE},
       adsurl = {https://ui.adsabs.harvard.edu/abs/2019MNRAS.485..889S}
}

@ARTICLE{Burrows2019,
       author = {{Burrows}, Adam and {Radice}, David and {Vartanyan}, David},
        title = "{Three-dimensional supernova explosion simulations of 9-, 10-, 11-, 12-, and 13-M$_{☉}$ stars}",
      journal = {\mnras},
         year = 2019,
        month = may,
       volume = {485},
       number = {3},
        pages = {3153-3168},
          doi = {10.1093/mnras/stz543},
archivePrefix = {arXiv},
       eprint = {1902.00547},
 primaryClass = {astro-ph.SR},
       adsurl = {https://ui.adsabs.harvard.edu/abs/2019MNRAS.485.3153B}
}

@ARTICLE{Vartanyan2019,
       author = {{Vartanyan}, David and {Burrows}, Adam and {Radice}, David and {Skinner}, M. Aaron and {Dolence}, Joshua},
        title = "{A successful 3D core-collapse supernova explosion model}",
      journal = {\mnras},
         year = 2019,
        month = jan,
       volume = {482},
       number = {1},
        pages = {351-369},
          doi = {10.1093/mnras/sty2585},
archivePrefix = {arXiv},
       eprint = {1809.05106},
 primaryClass = {astro-ph.HE},
       adsurl = {https://ui.adsabs.harvard.edu/abs/2019MNRAS.482..351V}
}

@article{Breen2013b,
  title = {Dynamical Evolution of Black Hole Subsystems in Idealized Star Clusters},
  author = {Breen, Philip G. and Heggie, Douglas C.},
  year = {2013},
  month = jul,
  journal = {Monthly Notices of the Royal Astronomical Society},
  volume = {432},
  number = {4},
  pages = {2779--2797},
  doi = {10.1093/mnras/stt628}
}

@ARTICLE{Dotter2026,
       author = {{Dotter}, Aaron and {Bauer}, Evan B. and {Park}, Minjung and {Conroy}, Charlie and {Milone}, Antonino P. and {Joyce}, Meridith and {Cantiello}, Matteo},
        title = "{MESA Isochrones and Stellar Tracks (MIST). II. Models with {\ensuremath{\alpha}}-enhanced Chemical Composition}",
      journal = {\apjs},
         year = 2026,
        month = apr,
       volume = {283},
       number = {2},
          eid = {64},
        pages = {64},
          doi = {10.3847/1538-4365/ae48f3},
archivePrefix = {arXiv},
       eprint = {2602.22012},
 primaryClass = {astro-ph.SR},
       adsurl = {https://ui.adsabs.harvard.edu/abs/2026ApJS..283...64D}
}

@ARTICLE{Bauer2026,
       author = {{Bauer}, Evan B. and {Dotter}, Aaron and {Conroy}, Charlie and {Cunningham}, Tim and {Park}, Minjung and {Tremblay}, Pier-Emmanuel},
        title = "{MESA Isochrones and Stellar Tracks (MIST). III. The White Dwarf Cooling Sequence}",
      journal = {\apjs},
         year = 2026,
        month = mar,
       volume = {283},
       number = {1},
          eid = {41},
        pages = {41},
          doi = {10.3847/1538-4365/ae401e},
archivePrefix = {arXiv},
       eprint = {2509.21717},
 primaryClass = {astro-ph.SR},
       adsurl = {https://ui.adsabs.harvard.edu/abs/2026ApJS..283...41B}
}

@article{Antonini2019,
  title = {Black Hole Growth through Hierarchical Black Hole Mergers in Dense Star Clusters: Implications for Gravitational Wave Detections},
  shorttitle = {Black Hole Growth through Hierarchical Black Hole Mergers in Dense Star Clusters},
  author = {Antonini, Fabio and Gieles, Mark and Gualandris, Alessia},
  year = 2019,
  month = jul,
  journal = {Monthly Notices of the Royal Astronomical Society},
  volume = {486},
  pages = {5008--5021},
  publisher = {OUP},
  doi = {10.1093/mnras/stz1149}
}

@ARTICLE{Antonini2020a,
       author = {{Antonini}, Fabio and {Gieles}, Mark},
        title = "{Population synthesis of black hole binary mergers from star clusters}",
      journal = {\mnras},
         year = 2020,
        month = feb,
       volume = {492},
       number = {2},
        pages = {2936-2954},
          doi = {10.1093/mnras/stz3584},
}

@ARTICLE{Antonini2020b,
       author = {{Antonini}, Fabio and {Gieles}, Mark},
        title = "{.Merger rate of black hole binaries from globular clusters: Theoretical error bars and comparison to gravitational wave data from GWTC-2}",
      journal = {\prd},
         year = 2020,
        month = dec,
       volume = {102},
       number = {12},
        pages = {123016},
          doi = {10.1103/PhysRevD.102.123016},
}

@article{RandoForastier2025,
  title = {Binary-Single Interactions with Different Mass Ratios: {{Implications}} for Gravitational Waves from Globular Clusters},
  shorttitle = {Binary-Single Interactions with Different Mass Ratios},
  author = {Rando Forastier, Bruno and Mar{\'i}n Pina, Daniel and Gieles, Mark and Portegies Zwart, Simon and Antonini, Fabio},
  year = 2025,
  month = may,
  journal = {Astronomy and Astrophysics},
  volume = {697},
  pages = {A118},
  publisher = {EDP},
  doi = {10.1051/0004-6361/202450890}
}

@ARTICLE{Antonini2023,
       author = {{Antonini}, Fabio and {Gieles}, Mark and {Dosopoulou}, Fani and {Chattopadhyay}, Debatri},
        title = "{Coalescing black hole binaries from globular clusters: mass distributions and comparison to gravitational wave data from GWTC-3}",
      journal = {\mnras},
         year = 2023,
        month = jun,
       volume = {522},
       number = {1},
        pages = {466-476},
          doi = {10.1093/mnras/stad972},
}

@ARTICLE{Watkins2015,
       author = {{Watkins}, Laura L. and {van der Marel}, Roeland P. and {Bellini}, Andrea and {Anderson}, Jay},
        title = "{Hubble Space Telescope Proper Motion (HSTPROMO) Catalogs of Galactic Globular Cluster. II. Kinematic Profiles and Maps}",
      journal = {\apj},
         year = 2015,
        month = apr,
       volume = {803},
       number = {1},
        pages = {29},
          doi = {10.1088/0004-637X/803/1/29},
}

@ARTICLE{Libralato2022,
       author = {{Libralato}, Mattia and {Bellini}, Andrea and {Vesperini}, Enrico and {Piotto}, Giampaolo and {Milone}, Antonino P. and {van der Marel}, Roeland P. and {Anderson}, Jay and {Aparicio}, Antonio and {Barbuy}, Beatriz and {Bedin}, Luigi R. and {Borsato}, Luca and {Cassisi}, Santi and {Dalessandro}, Emanuele and {Ferraro}, Francesco R. and {King}, Ivan R. and {Lanzoni}, Barbara and {Nardiello}, Domenico and {Ortolani}, Sergio and {Sarajedini}, Ata and {Sohn}, Sangmo Tony},
        title = "{The Hubble Space Telescope UV Legacy Survey of Galactic Globular Clusters. XXIII. Proper-motion Catalogs and Internal Kinematics}",
      journal = {\apj},
         year = 2022,
        month = aug,
       volume = {934},
       number = {2},
        pages = {150},
          doi = {10.3847/1538-4357/ac7727},
}

@ARTICLE{Lutzgendorf2013,
       author = {{L{\"u}tzgendorf}, N. and {Kissler-Patig}, M. and {Gebhardt}, K. and {Baumgardt}, H. and {Noyola}, E. and {de Zeeuw}, P.~T. and {Neumayer}, N. and {Jalali}, B. and {Feldmeier}, A.},
        title = "{Limits on intermediate-mass black holes in six Galactic globular clusters with integral-field spectroscopy}",
      journal = {\aap},
         year = 2013,
        month = apr,
       volume = {552},
          eid = {A49},
        pages = {A49},
          doi = {10.1051/0004-6361/201220307},
archivePrefix = {arXiv},
       eprint = {1212.3475},
 primaryClass = {astro-ph.GA},
       adsurl = {https://ui.adsabs.harvard.edu/abs/2013A&A...552A..49L}
}

@ARTICLE{Dalgleish2020,
       author = {{Dalgleish}, H. and {Kamann}, S. and {Usher}, C. and {Baumgardt}, H. and {Bastian}, N. and {Veitch-Michaelis}, J. and {Bellini}, A. and {Martocchia}, S. and {Da Costa}, G.~S. and {Mackey}, D. and {Bellstedt}, S. and {Pastorello}, N. and {Cerulo}, P.},
        title = "{The WAGGS project-III. Discrepant mass-to-light ratios of Galactic globular clusters at high metallicity}",
      journal = {\mnras},
         year = 2020,
        month = mar,
       volume = {492},
       number = {3},
        pages = {3859-3871},
          doi = {10.1093/mnras/staa091},
}

@ARTICLE{Kamann2018,
       author = {{Kamann}, S. and {Husser}, T. -O. and {Dreizler}, S. and {Emsellem}, E. and {Weilbacher}, P.~M. and {Martens}, S. and {Bacon}, R. and {den Brok}, M. and {Giesers}, B. and {Krajnovi{\'c}}, D. and {Roth}, M.~M. and {Wendt}, M. and {Wisotzki}, L.},
        title = "{A stellar census in globular clusters with MUSE: The contribution of rotation to cluster dynamics studied with 200 000 stars}",
      journal = {\mnras},
         year = 2018,
        month = feb,
       volume = {473},
       number = {4},
        pages = {5591-5616},
          doi = {10.1093/mnras/stx2719},
}

@ARTICLE{deBoer2019,
       author = {{de Boer}, T.~J.~L. and {Gieles}, M. and {Balbinot}, E. and {H{\'e}nault-Brunet}, V. and {Sollima}, A. and {Watkins}, L.~L. and {Claydon}, I.},
        title = "{Globular cluster number density profiles using Gaia DR2}",
      journal = {\mnras},
         year = 2019,
        month = jun,
       volume = {485},
       number = {4},
        pages = {4906-4935},
          doi = {10.1093/mnras/stz651},
}

@ARTICLE{Miocchi2013,
       author = {{Miocchi}, P. and {Lanzoni}, B. and {Ferraro}, F.~R. and {Dalessandro}, E. and {Vesperini}, E. and {Pasquato}, M. and {Beccari}, G. and {Pallanca}, C. and {Sanna}, N.},
        title = "{Star Count Density Profiles and Structural Parameters of 26 Galactic Globular Clusters}",
      journal = {\apj},
         year = 2013,
        month = sep,
       volume = {774},
       number = {2},
        pages = {151},
          doi = {10.1088/0004-637X/774/2/151},
}

@ARTICLE{Speagle2020,
       author = {{Speagle}, Joshua S.},
        title = "{DYNESTY: a dynamic nested sampling package for estimating Bayesian posteriors and evidences}",
      journal = {\mnras},
         year = 2020,
        month = apr,
       volume = {493},
       number = {3},
        pages = {3132-3158},
          doi = {10.1093/mnras/staa278},
}

@ARTICLE{Giesers2018,
       author = {{Giesers}, Benjamin and {Dreizler}, Stefan and {Husser}, Tim-Oliver and {Kamann}, Sebastian and {Anglada Escud{\'e}}, Guillem and {Brinchmann}, Jarle and {Carollo}, C. Marcella and {Roth}, Martin M. and {Weilbacher}, Peter M. and {Wisotzki}, Lutz},
        title = "{A detached stellar-mass black hole candidate in the globular cluster NGC 3201}",
      journal = {\mnras},
         year = 2018,
        month = mar,
       volume = {475},
       number = {1},
        pages = {L15-L19},
          doi = {10.1093/mnrasl/slx203},
}

@ARTICLE{Giesers2019,
       author = {{Giesers}, Benjamin and {Kamann}, Sebastian and {Dreizler}, Stefan and {Husser}, Tim-Oliver and {Askar}, Abbas and {G{\"o}ttgens}, Fabian and {Brinchmann}, Jarle and {Latour}, Marilyn and {Weilbacher}, Peter M. and {Wendt}, Martin and {Roth}, Martin M.},
        title = "{A stellar census in globular clusters with MUSE: Binaries in NGC 3201}",
      journal = {\aap},
         year = 2019,
        month = dec,
       volume = {632},
        pages = {A3},
          doi = {10.1051/0004-6361/201936203},
}

@ARTICLE{Whitaker2026,
       author = {{Whitaker}, Matthew and {Kerr}, Evan and {Seth}, Anil and {H{\"a}berle}, Maximilian and {Strader}, Jay and {Anderson}, Jay and {Bellini}, Andrea and {Clontz}, Callie and {Freeman}, Zack and {Griggio}, Massimo and {Kamann}, Sebastian and {Libralato}, Mattia and {Neumayer}, Nadine and {Gonz{\'a}lez Prieto}, Elena and {Rodriguez}, Carl L. and {Saracino}, Sara and {Smith}, Peter and {van de Ven}, Glenn and {Wang}, Zixian},
        title = "{A Long Period Stellar-mass Black Hole Binary in {\ensuremath{\omega}} Centauri}",
      journal = {\apjl},
         year = 2026,
        month = jul,
       volume = {1006},
       number = {1},
          eid = {L1},
        pages = {L1},
          doi = {10.3847/2041-8213/ae7a5c},
archivePrefix = {arXiv},
       eprint = {2606.18350},
 primaryClass = {astro-ph.GA},
       adsurl = {https://ui.adsabs.harvard.edu/abs/2026ApJ..1006L...1W}
}

@ARTICLE{Haberle2025,
       author = {{H{\"a}berle}, Maximilian and {Neumayer}, N. and {Clontz}, C. and {Seth}, A.~C. and {Smith}, P.~J. and {Kamann}, S. and {Pechetti}, R. and {Nitschai}, M.~S. and {Alfaro-Cuello}, M. and {Baumgardt}, H. and {Bellini}, A. and {Feldmeier-Krause}, A. and {Kacharov}, N. and {Libralato}, M. and {Milone}, A.~P. and {Souza}, S.~O. and {van de Ven}, G. and {Wang}, Z.},
        title = "{oMEGACat. VI. Analysis of the Overall Kinematics of Omega Centauri in 3D: Velocity Dispersion, Kinematic Distance, Anisotropy, and Energy Equipartition}",
      journal = {\apj},
         year = 2025,
        month = apr,
       volume = {983},
       number = {2},
          eid = {95},
        pages = {95},
          doi = {10.3847/1538-4357/adbe67},
archivePrefix = {arXiv},
       eprint = {2503.04903},
 primaryClass = {astro-ph.GA},
       adsurl = {https://ui.adsabs.harvard.edu/abs/2025ApJ...983...95H}
}

@article{Giersz2013,
  title = {{{MOCCA}} Code for Star Cluster Simulations -- {{II}}. {{Comparison}} with {{N-body}} Simulations},
  author = {Giersz, Mirek and Heggie, Douglas C. and Hurley, Jarrod R. and Hypki, Arkadiusz},
  year = 2013,
  month = may,
  journal = {Monthly Notices of the Royal Astronomical Society},
  volume = {431},
  number = {3},
  pages = {2184--2199},
  doi = {10.1093/mnras/stt307}
}

@article{Hypki2013,
  title = {Mocca Code for Star Cluster Simulations -- {{I}}. {{Blue}} Stragglers, First Results},
  author = {Hypki, Arkadiusz and Giersz, Mirek},
  year = 2013,
  month = feb,
  journal = {Monthly Notices of the Royal Astronomical Society},
  volume = {429},
  number = {2},
  pages = {1221--1243},
  doi = {10.1093/mnras/sts415}
}

@article{Adamo2024,
  title = {Bound Star Clusters Observed in a Lensed Galaxy 460 {{Myr}} after the {{Big Bang}}},
  author = {Adamo, Angela and Bradley, Larry D. and Vanzella, Eros and Claeyssens, Ad{\'e}la{\"i}de and Welch, Brian and Diego, Jose M. and Mahler, Guillaume and Oguri, Masamune and Sharon, Keren and {Abdurro'uf} and Hsiao, Tiger Yu-Yang and Xu, Xinfeng and Messa, Matteo and Lassen, Augusto E. and Zackrisson, Erik and Brammer, Gabriel and Coe, Dan and Kokorev, Vasily and Ricotti, Massimo and Zitrin, Adi and Fujimoto, Seiji and Inoue, Akio K. and Resseguier, Tom and Rigby, Jane R. and {Jim{\'e}nez-Teja}, Yolanda and Windhorst, Rogier A. and Hashimoto, Takuya and Tamura, Yoichi},
  year = 2024,
  month = aug,
  journal = {Nature},
  volume = {632},
  pages = {513--516},
  doi = {10.1038/s41586-024-07703-7}
}

@article{Vanzella2023,
  title = {{{JWST}}/{{NIRCam Probes Young Star Clusters}} in the {{Reionization Era Sunrise Arc}}},
  author = {Vanzella, Eros and Claeyssens, Ad{\'e}la{\"i}de and Welch, Brian and Adamo, Angela and Coe, Dan and Diego, Jose M. and Mahler, Guillaume and Khullar, Gourav and Kokorev, Vasily and Oguri, Masamune and Ravindranath, Swara and Furtak, Lukas J. and Hsiao, Tiger Yu-Yang and {Abdurro'uf} and Mandelker, Nir and Brammer, Gabriel and Bradley, Larry D. and Brada{\v c}, Maru{\v s}a and Conselice, Christopher J. and Dayal, Pratika and Nonino, Mario and {Andrade-Santos}, Felipe and Windhorst, Rogier A. and Pirzkal, Nor and Sharon, Keren and {de Mink}, S. E. and Fujimoto, Seiji and Zitrin, Adi and Eldridge, Jan J. and Norman, Colin},
  year = 2023,
  month = mar,
  journal = {The Astrophysical Journal},
  volume = {945},
  pages = {53},
  publisher = {IOP},
  doi = {10.3847/1538-4357/acb59a}
}

@article{Claeyssens2023,
  title = {Star Formation at the Smallest Scales: A {{JWST}} Study of the Clump Populations in {{SMACS0723}}},
  shorttitle = {Star Formation at the Smallest Scales},
  author = {Claeyssens, Ad{\'e}la{\"i}de and Adamo, Angela and Richard, Johan and Mahler, Guillaume and Messa, Matteo and {Dessauges-Zavadsky}, Miroslava},
  year = 2023,
  month = apr,
  journal = {Monthly Notices of the Royal Astronomical Society},
  volume = {520},
  pages = {2180--2203},
  publisher = {OUP},
  doi = {10.1093/mnras/stac3791}
}

@misc{Claeyssens2026,
  title = {A First {{GLIMPSE}} into Star Clusters Populations across Cosmic Time},
  author = {Claeyssens, Ad{\'e}la{\"i}de and Adamo, Angela and Kokorev, Vasily and Furtak, Lukas and Richard, Johan and Beauchesne, Benjamin and {Dessauges-Zavadsky}, Miroslava and Atek, Hakim and Chisholm, John and Endsley, Ryan and Fujimoto, Seiji and Korber, Damien and Pan, Richard and {Saldana-Lopez}, Alberto and Schaerer, Daniel},
  year = 2026,
  month = jan,
  publisher = {arXiv},
  doi = {10.48550/arXiv.2601.16281}
}

@ARTICLE{Mandel2016,
       author = {{Mandel}, Ilya},
        title = "{Estimates of black hole natal kick velocities from observations of low-mass X-ray binaries}",
      journal = {\mnras},
         year = 2016,
        month = feb,
       volume = {456},
       number = {1},
        pages = {578-581},
          doi = {10.1093/mnras/stv2733},
archivePrefix = {arXiv},
       eprint = {1510.03871},
 primaryClass = {astro-ph.HE},
       adsurl = {https://ui.adsabs.harvard.edu/abs/2016MNRAS.456..578M}
}

@article{Repetto2017,
  title = {The {{Galactic}} Distribution of {{X-ray}} Binaries and Its Implications for Compact Object Formation and Natal Kicks},
  author = {Repetto, Serena and Igoshev, Andrei P. and Nelemans, Gijs},
  year = 2017,
  month = may,
  journal = {Monthly Notices of the Royal Astronomical Society},
  volume = {467},
  pages = {298--310},
  publisher = {OUP},
  doi = {10.1093/mnras/stx027}
}

@article{Atri2019,
  title = {Potential Kick Velocity Distribution of Black Hole {{X-ray}} Binaries and Implications for Natal Kicks},
  author = {Atri, P. and {Miller-Jones}, J. C. A. and Bahramian, A. and Plotkin, R. M. and Jonker, P. G. and Nelemans, G. and Maccarone, T. J. and Sivakoff, G. R. and Deller, A. T. and Chaty, S. and Torres, M. A. P. and Horiuchi, S. and McCallum, J. and Natusch, T. and Phillips, C. J. and Stevens, J. and Weston, S.},
  year = 2019,
  month = nov,
  journal = {Monthly Notices of the Royal Astronomical Society},
  volume = {489},
  pages = {3116--3134},
  publisher = {OUP},
  doi = {10.1093/mnras/stz2335}
}

@ARTICLE{Popov2025,
       author = {{Popov}, Sergei and {M{\"u}ller}, Bernhard and {Mandel}, Ilya},
        title = "{Natal kicks of compact objects}",
      journal = {\nar},
         year = 2025,
        month = dec,
       volume = {101},
          eid = {101734},
        pages = {101734},
          doi = {10.1016/j.newar.2025.101734},
archivePrefix = {arXiv},
       eprint = {2509.01430},
 primaryClass = {astro-ph.HE},
       adsurl = {https://ui.adsabs.harvard.edu/abs/2025NewAR.10101734P}
}

@ARTICLE{Janka1994,
       author = {{Janka}, H.-T. and {Mueller}, E.},
        title = "{Neutron star recoils from anisotropic supernovae.}",
      journal = {\aap},
         year = 1994,
        month = oct,
       volume = {290},
        pages = {496-502},
       adsurl = {https://ui.adsabs.harvard.edu/abs/1994A&A...290..496J}
}

@ARTICLE{Burrows1996,
       author = {{Burrows}, Adam and {Hayes}, John},
        title = "{Pulsar Recoil and Gravitational Radiation Due to Asymmetrical Stellar Collapse and Explosion}",
      journal = {\prl},
         year = 1996,
        month = jan,
       volume = {76},
       number = {3},
        pages = {352-355},
          doi = {10.1103/PhysRevLett.76.352},
archivePrefix = {arXiv},
       eprint = {astro-ph/9511106},
 primaryClass = {astro-ph},
       adsurl = {https://ui.adsabs.harvard.edu/abs/1996PhRvL..76..352B}
}

@ARTICLE{Burrows2021,
       author = {{Burrows}, A. and {Vartanyan}, D.},
        title = "{Core-collapse supernova explosion theory}",
      journal = {\nat},
         year = 2021,
        month = jan,
       volume = {589},
       number = {7840},
        pages = {29-39},
          doi = {10.1038/s41586-020-03059-w},
archivePrefix = {arXiv},
       eprint = {2009.14157},
 primaryClass = {astro-ph.SR},
       adsurl = {https://ui.adsabs.harvard.edu/abs/2021Natur.589...29B}
}

@ARTICLE{Willcox2025a,
       author = {{Willcox}, R. and {Marchant}, P. and {Vigna-G{\'o}mez}, A. and {Sana}, H. and {Bodensteiner}, J. and {Deshmukh}, K. and {Esseldeurs}, M. and {Fabry}, M. and {H{\'e}nault-Brunet}, V. and {Janssens}, S. and {Mahy}, L. and {Patrick}, L. and {Pauli}, D. and {Renzo}, M. and {Sander}, A.~A.~C. and {Shenar}, T. and {van Son}, L.~A.~C. and {Stoop}, M.},
        title = "{Binarity at LOw Metallicity (BLOeM): Bayesian inference of natal kicks from inert black hole binaries}",
      journal = {\aap},
         year = 2025,
        month = aug,
       volume = {700},
          eid = {A59},
        pages = {A59},
          doi = {10.1051/0004-6361/202555274},
archivePrefix = {arXiv},
       eprint = {2504.16669},
 primaryClass = {astro-ph.SR},
       adsurl = {https://ui.adsabs.harvard.edu/abs/2025A&A...700A..59W}
}

@ARTICLE{Willcox2025b,
       author = {{Willcox}, Reinhold and {Schneider}, Fabian R.~N. and {Laplace}, Eva and {Podsiadlowski}, Philipp and {Maltsev}, Kiril and {Mandel}, Ilya and {Marchant}, Pablo and {Sana}, Hugues and {Li}, Tjonnie G.~F. and {Hertog}, Thomas},
        title = "{Good things always come in 3s: trimodality in the binary black-hole chirp-mass distribution supports bimodal black-hole formation}",
      journal = {arXiv e-prints},
         year = 2025,
        month = oct,
          eid = {arXiv:2510.07573},
        pages = {arXiv:2510.07573},
          doi = {10.48550/arXiv.2510.07573},
archivePrefix = {arXiv},
       eprint = {2510.07573},
 primaryClass = {astro-ph.SR},
       adsurl = {https://ui.adsabs.harvard.edu/abs/2025arXiv251007573W}
}

@ARTICLE{Chan2018,
       author = {{Chan}, Conrad and {M{\"u}ller}, Bernhard and {Heger}, Alexander and {Pakmor}, R{\"u}diger and {Springel}, Volker},
        title = "{Black Hole Formation and Fallback during the Supernova Explosion of a 40 M $_{☉}$ Star}",
      journal = {\apjl},
         year = 2018,
        month = jan,
       volume = {852},
       number = {1},
          eid = {L19},
        pages = {L19},
          doi = {10.3847/2041-8213/aaa28c},
archivePrefix = {arXiv},
       eprint = {1710.00838},
 primaryClass = {astro-ph.SR},
       adsurl = {https://ui.adsabs.harvard.edu/abs/2018ApJ...852L..19C}
}

@ARTICLE{Chan2020,
       author = {{Chan}, Conrad and {M{\"u}ller}, Bernhard and {Heger}, Alexander},
        title = "{The impact of fallback on the compact remnants and chemical yields of core-collapse supernovae}",
      journal = {\mnras},
         year = 2020,
        month = jul,
       volume = {495},
       number = {4},
        pages = {3751-3762},
          doi = {10.1093/mnras/staa1431},
archivePrefix = {arXiv},
       eprint = {2003.04320},
 primaryClass = {astro-ph.SR},
       adsurl = {https://ui.adsabs.harvard.edu/abs/2020MNRAS.495.3751C}
}

@ARTICLE{Janka2024,
       author = {{Janka}, Hans-Thomas and {Kresse}, Daniel},
        title = "{Interplay between neutrino kicks and hydrodynamic kicks of neutron stars and black holes}",
      journal = {\apss},
         year = 2024,
        month = aug,
       volume = {369},
       number = {8},
          eid = {80},
        pages = {80},
          doi = {10.1007/s10509-024-04343-1},
archivePrefix = {arXiv},
       eprint = {2401.13817},
 primaryClass = {astro-ph.HE},
       adsurl = {https://ui.adsabs.harvard.edu/abs/2024Ap&SS.369...80J}
}

@ARTICLE{Bocchi2026,
       author = {{Bocchi}, Viola and {Liempi}, Mat{\'\i}as and {Schleicher}, Dominik R.~G.},
        title = "{Formation of intermediate-mass black holes in young massive clusters detected with JWST: analytic mass estimates}",
      journal = {arXiv e-prints},
         year = 2026,
        month = may,
          eid = {arXiv:2605.20381},
        pages = {arXiv:2605.20381},
          doi = {10.48550/arXiv.2605.20381},
archivePrefix = {arXiv},
       eprint = {2605.20381},
 primaryClass = {astro-ph.GA},
       adsurl = {https://ui.adsabs.harvard.edu/abs/2026arXiv260520381B}
}

@ARTICLE{Vergara2026,
       author = {{Vergara}, M.~C. and {Askar}, A. and {Flammini Dotti}, F. and {Schleicher}, D.~R.~G. and {Escala}, A. and {Spurzem}, R. and {Giersz}, M. and {Hurley}, J. and {Arca Sedda}, M. and {Neumayer}, N.},
        title = "{Efficient black hole seed formation in low-metallicity and dense stellar clusters with implications for JWST sources}",
      journal = {\aap},
         year = 2026,
        month = mar,
       volume = {707},
          eid = {A71},
        pages = {A71},
          doi = {10.1051/0004-6361/202556878},
archivePrefix = {arXiv},
       eprint = {2508.14260},
 primaryClass = {astro-ph.GA},
       adsurl = {https://ui.adsabs.harvard.edu/abs/2026A&A...707A..71V}
}

@article{Rantala2026,
  title = {{{FROST-CLUSTERS}} - {{III}}. {{Metallicity-dependent}} Intermediate-Mass Black Hole Formation by Runaway Collisions in Dense Star Clusters},
  author = {Rantala, Antti and Naab, Thorsten and Lah{\'e}n, Natalia and Reuter, Klaus and Rampp, Markus and Chru{\'s}li{\'n}ska, Martyna and Reinoso, Basti{\'a}n},
  year = 2026,
  month = jul,
  journal = {Monthly Notices of the Royal Astronomical Society},
  volume = {549},
  pages = {stag986},
  publisher = {OUP},
  doi = {10.1093/mnras/stag986}
}

@ARTICLE{Hunter2007,
       author = {{Hunter}, John D.},
        title = "{Matplotlib: A 2D Graphics Environment}",
      journal = {Computing in Science and Engineering},
         year = 2007,
        month = may,
       volume = {9},
       number = {3},
        pages = {90-95},
          doi = {10.1109/MCSE.2007.55},
}

@ARTICLE{Astropy2018,
       author = {{Astropy Collaboration} and {Price-Whelan}, A.~M. and {Sip{\H{o}}cz}, B.~M. and {G{\"u}nther}, H.~M. and {Lim}, P.~L. and {Crawford}, S.~M. and {Conseil}, S. and {Shupe}, D.~L. and {Craig}, M.~W. and {Dencheva}, N. and {Ginsburg}, A. and {VanderPlas}, J.~T. and {Bradley}, L.~D. and {P{\'e}rez-Su{\'a}rez}, D. and {de Val-Borro}, M. and {Aldcroft}, T.~L. and {Cruz}, K.~L. and {Robitaille}, T.~P. and {Tollerud}, E.~J. and {Ardelean}, C. and {Babej}, T. and {Bach}, Y.~P. and {Bachetti}, M. and {Bakanov}, A.~V. and {Bamford}, S.~P. and {Barentsen}, G. and {Barmby}, P. and {Baumbach}, A. and {Berry}, K.~L. and {Biscani}, F. and {Boquien}, M. and {Bostroem}, K.~A. and {Bouma}, L.~G. and {Brammer}, G.~B. and {Bray}, E.~M. and {Breytenbach}, H. and {Buddelmeijer}, H. and {Burke}, D.~J. and {Calderone}, G. and {Cano Rodr{\'\i}guez}, J.~L. and {Cara}, M. and {Cardoso}, J.~V.~M. and {Cheedella}, S. and {Copin}, Y. and {Corrales}, L. and {Crichton}, D. and {D'Avella}, D. and {Deil}, C. and {Depagne}, {\'E}. and {Dietrich}, J.~P. and {Donath}, A. and {Droettboom}, M. and {Earl}, N. and {Erben}, T. and {Fabbro}, S. and {Ferreira}, L.~A. and {Finethy}, T. and {Fox}, R.~T. and {Garrison}, L.~H. and {Gibbons}, S.~L.~J. and {Goldstein}, D.~A. and {Gommers}, R. and {Greco}, J.~P. and {Greenfield}, P. and {Groener}, A.~M. and {Grollier}, F. and {Hagen}, A. and {Hirst}, P. and {Homeier}, D. and {Horton}, A.~J. and {Hosseinzadeh}, G. and {Hu}, L. and {Hunkeler}, J.~S. and {Ivezi{\'c}}, {\v{Z}}. and {Jain}, A. and {Jenness}, T. and {Kanarek}, G. and {Kendrew}, S. and {Kern}, N.~S. and {Kerzendorf}, W.~E. and {Khvalko}, A. and {King}, J. and {Kirkby}, D. and {Kulkarni}, A.~M. and {Kumar}, A. and {Lee}, A. and {Lenz}, D. and {Littlefair}, S.~P. and {Ma}, Z. and {Macleod}, D.~M. and {Mastropietro}, M. and {McCully}, C. and {Montagnac}, S. and {Morris}, B.~M. and {Mueller}, M. and {Mumford}, S.~J. and {Muna}, D. and {Murphy}, N.~A. and {Nelson}, S. and {Nguyen}, G.~H. and {Ninan}, J.~P. and {N{\"o}the}, M. and {Ogaz}, S. and {Oh}, S. and {Parejko}, J.~K. and {Parley}, N. and {Pascual}, S. and {Patil}, R. and {Patil}, A.~A. and {Plunkett}, A.~L. and {Prochaska}, J.~X. and {Rastogi}, T. and {Reddy Janga}, V. and {Sabater}, J. and {Sakurikar}, P. and {Seifert}, M. and {Sherbert}, L.~E. and {Sherwood-Taylor}, H. and {Shih}, A.~Y. and {Sick}, J. and {Silbiger}, M.~T. and {Singanamalla}, S. and {Singer}, L.~P. and {Sladen}, P.~H. and {Sooley}, K.~A. and {Sornarajah}, S. and {Streicher}, O. and {Teuben}, P. and {Thomas}, S.~W. and {Tremblay}, G.~R. and {Turner}, J.~E.~H. and {Terr{\'o}n}, V. and {van Kerkwijk}, M.~H. and {de la Vega}, A. and {Watkins}, L.~L. and {Weaver}, B.~A. and {Whitmore}, J.~B. and {Woillez}, J. and {Zabalza}, V. and {Astropy Contributors}},
        title = "{The Astropy Project: Building an Open-science Project and Status of the v2.0 Core Package}",
      journal = {\aj},
         year = 2018,
        month = sep,
       volume = {156},
       number = {3},
        pages = {123},
          doi = {10.3847/1538-3881/aabc4f},
}

@ARTICLE{Harris2020,
       author = {{Harris}, Charles R. and {Millman}, K. Jarrod and {van der Walt}, St{\'e}fan J. and {Gommers}, Ralf and {Virtanen}, Pauli and {Cournapeau}, David and {Wieser}, Eric and {Taylor}, Julian and {Berg}, Sebastian and {Smith}, Nathaniel J. and {Kern}, Robert and {Picus}, Matti and {Hoyer}, Stephan and {van Kerkwijk}, Marten H. and {Brett}, Matthew and {Haldane}, Allan and {del R{\'\i}o}, Jaime Fern{\'a}ndez and {Wiebe}, Mark and {Peterson}, Pearu and {G{\'e}rard-Marchant}, Pierre and {Sheppard}, Kevin and {Reddy}, Tyler and {Weckesser}, Warren and {Abbasi}, Hameer and {Gohlke}, Christoph and {Oliphant}, Travis E.},
        title = "{Array programming with NumPy}",
      journal = {\nat},
         year = 2020,
        month = sep,
       volume = {585},
       number = {7825},
        pages = {357-362},
          doi = {10.1038/s41586-020-2649-2},
}

@ARTICLE{Virtanen2020,
       author = {{Virtanen}, Pauli and {Gommers}, Ralf and {Oliphant}, Travis E. and {Haberland}, Matt and {Reddy}, Tyler and {Cournapeau}, David and {Burovski}, Evgeni and {Peterson}, Pearu and {Weckesser}, Warren and {Bright}, Jonathan and {van der Walt}, St{\'e}fan J. and {Brett}, Matthew and {Wilson}, Joshua and {Millman}, K. Jarrod and {Mayorov}, Nikolay and {Nelson}, Andrew R.~J. and {Jones}, Eric and {Kern}, Robert and {Larson}, Eric and {Carey}, C.~J. and {Polat}, {\.I}lhan and {Feng}, Yu and {Moore}, Eric W. and {VanderPlas}, Jake and {Laxalde}, Denis and {Perktold}, Josef and {Cimrman}, Robert and {Henriksen}, Ian and {Quintero}, E.~A. and {Harris}, Charles R. and {Archibald}, Anne M. and {Ribeiro}, Ant{\^o}nio H. and {Pedregosa}, Fabian and {van Mulbregt}, Paul and {SciPy 1. 0 Contributors}},
        title = "{SciPy 1.0: fundamental algorithms for scientific computing in Python}",
      journal = {Nature Methods},
         year = 2020,
        month = feb,
       volume = {17},
        pages = {261-272},
          doi = {10.1038/s41592-019-0686-2},
archivePrefix = {arXiv},
       eprint = {1907.10121},
 primaryClass = {cs.MS},
       adsurl = {https://ui.adsabs.harvard.edu/abs/2020NatMe..17..261V}
}

@misc{Collette2022,
  author       = {Andrew Collette and Thomas Kluyver and Thomas A Caswell and James Tocknell and Jerome Kieffer and Aleksandar Jelenak and Anthony Scopatz and Darren Dale and Chen and Thomas VINCENT and Matt Einhorn and payno and juliagarriga and Pierlauro Sciarelli and Valentin Valls and Satrajit Ghosh and Ulrik Kofoed Pedersen and jakirkham and Martin Raspaud and Cyril Danilevski and Hameer Abbasi and John Readey and Kai Mühlbauer and Andrey Paramonov and Lawrence Chan and V. Armando Solé and jialin and Daniel Hay Guest and Yu Feng and Mar Kittisopikul},
  title        = {h5py: 3.7.0},
  month        = may,
  year         = 2022,
  publisher    = {Zenodo},
  version      = {3.7.0},
  doi          = {10.5281/zenodo.6575970},
}

@misc{Koposov2025,
  author       = {Sergey Koposov and Josh Speagle and Kyle Barbary and Gregory Ashton and Ed Bennett and Johannes Buchner and Carl Scheffler and Colm Talbot and Ben Cook and James Guillochon and Patricio Cubillos and Andrés Asensio Ramos and Matthieu Dartiailh and Ilya and Erik Tollerud and Dustin Lang and Ben Johnson and jtmendel and Edward Higson and Thomas Vandal and Tansu Daylan and Ruth Angus and patelR and Phillip Cargile and Patrick Sheehan and Matt Pitkin and Matthew Kirk and Lu Xu and Joel Leja and joezuntz},
  title        = {joshspeagle/dynesty: v3.0.0},
  month        = oct,
  year         = 2025,
  publisher    = {Zenodo},
  version      = {v3.0.0},
  doi          = {10.5281/zenodo.17268284},
  url          = {https://doi.org/10.5281/zenodo.17268284},
}

@misc{cmctoolkit,
         author = {Nicholas Z. Rui and Kyle Kremer and Newlin C. Weatherford and Sourav Chatterjee and Frederic A. Rasio and Carl L. Rodriguez and Claire S. Ye},
          title = {NicholasRui/cmctoolkit: First release},
          month = mar,
           year = 2021,
      publisher = {Zenodo},
        version = {1.0},
            doi = {10.5281/zenodo.4579951},
            url = {https://doi.org/10.5281/zenodo.4579951}
}

@article{PortegiesZwart2010,
  title = {Young {{Massive Star Clusters}}},
  author = {Portegies Zwart, Simon F. and McMillan, Stephen L. W. and Gieles, Mark},
  year = 2010,
  month = sep,
  journal = {Annual Review of Astronomy and Astrophysics},
  volume = {48},
  pages = {431--493},
  doi = {10.1146/annurev-astro-081309-130834}
}

@ARTICLE{Cournoyer-Cloutier2024,
       author = {{Cournoyer-Cloutier}, Claude and {Sills}, Alison and {Harris}, William E. and {Polak}, Brooke and {Rieder}, Steven and {Andersson}, Eric P. and {Appel}, Sabrina M. and {Mac Low}, Mordecai-Mark and {McMillan}, Stephen and {Portegies Zwart}, Simon},
        title = "{Massive Star Cluster Formation with Binaries. I. Evolution of Binary Populations}",
      journal = {\apj},
         year = 2024,
        month = dec,
       volume = {977},
       number = {2},
          eid = {203},
        pages = {203},
          doi = {10.3847/1538-4357/ad90b3},
archivePrefix = {arXiv},
       eprint = {2410.07433},
 primaryClass = {astro-ph.GA},
       adsurl = {https://ui.adsabs.harvard.edu/abs/2024ApJ...977..203C}
}

@ARTICLE{Polak2024,
       author = {{Polak}, Brooke and {Mac Low}, Mordecai-Mark and {Klessen}, Ralf S. and {Wei Teh}, Jia and {Cournoyer-Cloutier}, Claude and {Andersson}, Eric P. and {Appel}, Sabrina M. and {Tran}, Aaron and {Lewis}, Sean C. and {Wilhelm}, Maite J.~C. and {Portegies Zwart}, Simon and {Glover}, Simon C.~O. and {Rieder}, Steven and {Wang}, Long and {McMillan}, Stephen L.~W.},
        title = "{Massive star cluster formation: I. High star formation efficiency while resolving feedback of individual stars}",
      journal = {\aap},
         year = 2024,
        month = oct,
       volume = {690},
          eid = {A94},
        pages = {A94},
          doi = {10.1051/0004-6361/202348840},
archivePrefix = {arXiv},
       eprint = {2312.06509},
 primaryClass = {astro-ph.GA},
       adsurl = {https://ui.adsabs.harvard.edu/abs/2024A&A...690A..94P}
}

@ARTICLE{Reina-Campos2025,
       author = {{Reina-Campos}, Marta and {Gnedin}, Oleg Y. and {Sills}, Alison and {Li}, Hui},
        title = "{The Star Clusters as Links between Galaxy Evolution and Star Formation Project. I. Numerical Method}",
      journal = {\apj},
         year = 2025,
        month = jan,
       volume = {978},
       number = {1},
          eid = {15},
        pages = {15},
          doi = {10.3847/1538-4357/ad909f},
archivePrefix = {arXiv},
       eprint = {2408.04694},
 primaryClass = {astro-ph.GA},
       adsurl = {https://ui.adsabs.harvard.edu/abs/2025ApJ...978...15R}
}

@article{Lahen2025a,
  title = {The Formation, Evolution, and Disruption of Star Clusters with Improved Gravitational Dynamics in Simulated Dwarf Galaxies},
  author = {Lah{\'e}n, Natalia and Rantala, Antti and Naab, Thorsten and Partmann, Christian and Johansson, Peter H. and Hislop, Jessica May},
  year = 2025,
  month = apr,
  journal = {Monthly Notices of the Royal Astronomical Society},
  volume = {538},
  pages = {2129--2148},
  publisher = {OUP},
  doi = {10.1093/mnras/staf350}
}

@article{Lahen2025b,
  title = {Mergers All the Way down: Stellar Collisions and Kinematics of a Dense Hierarchically Forming Massive Star Cluster in a Dwarf Starburst},
  shorttitle = {Mergers All the Way Down},
  author = {Lah{\'e}n, Natalia and Naab, Thorsten and Rantala, Antti and Partmann, Christian},
  year = 2025,
  month = oct,
  journal = {Monthly Notices of the Royal Astronomical Society},
  volume = {543},
  pages = {1023--1038},
  publisher = {OUP},
  doi = {10.1093/mnras/staf1546}
}

@ARTICLE{Williams2025,
       author = {{Williams}, Claire E. and {Naoz}, Smadar and {Lake}, William and {Burkhart}, Blakesley and {Marinacci}, Federico and {Vogelsberger}, Mark and {Yoshida}, Naoki and {Menon}, Shyam H. and {Chen}, Avi and {Adamo}, Angela},
        title = "{{\ensuremath{\Lambda}}CDM Star Clusters at Cosmic Dawn: Stellar Densities, Environment, and Equilibrium}",
      journal = {\apj},
         year = 2025,
        month = sep,
       volume = {990},
       number = {2},
          eid = {135},
        pages = {135},
          doi = {10.3847/1538-4357/adf19d},
archivePrefix = {arXiv},
       eprint = {2502.17561},
 primaryClass = {astro-ph.GA},
       adsurl = {https://ui.adsabs.harvard.edu/abs/2025ApJ...990..135W}
}

@ARTICLE{LIGO2026,
       author = {{The LIGO Scientific Collaboration} and {the Virgo Collaboration} and {the KAGRA Collaboration}},
        title = "{GWTC-5.0: Population Properties of Merging Compact Binaries}",
      journal = {arXiv e-prints},
         year = 2026,
        month = may,
          eid = {arXiv:2605.27226},
        pages = {arXiv:2605.27226},
          doi = {10.48550/arXiv.2605.27226},
archivePrefix = {arXiv},
       eprint = {2605.27226},
 primaryClass = {astro-ph.HE},
       adsurl = {https://ui.adsabs.harvard.edu/abs/2026arXiv260527226T}
}

@ARTICLE{Wang2020,
       author = {{Wang}, Long and {Iwasawa}, Masaki and {Nitadori}, Keigo and {Makino}, Junichiro},
        title = "{PETAR: a high-performance N-body code for modelling massive collisional stellar systems}",
      journal = {\mnras},
         year = 2020,
        month = sep,
       volume = {497},
       number = {1},
        pages = {536-555},
          doi = {10.1093/mnras/staa1915},
archivePrefix = {arXiv},
       eprint = {2006.16560},
 primaryClass = {astro-ph.IM},
       adsurl = {https://ui.adsabs.harvard.edu/abs/2020MNRAS.497..536W}
}

@ARTICLE{Mapelli2020,
       author = {{Mapelli}, Michela and {Spera}, Mario and {Montanari}, Enrico and {Limongi}, Marco and {Chieffi}, Alessandro and {Giacobbo}, Nicola and {Bressan}, Alessandro and {Bouffanais}, Yann},
        title = "{Impact of the Rotation and Compactness of Progenitors on the Mass of Black Holes}",
      journal = {\apj},
         year = 2020,
        month = jan,
       volume = {888},
       number = {2},
          eid = {76},
        pages = {76},
          doi = {10.3847/1538-4357/ab584d},
archivePrefix = {arXiv},
       eprint = {1909.01371},
 primaryClass = {astro-ph.HE},
       adsurl = {https://ui.adsabs.harvard.edu/abs/2020ApJ...888...76M}
}

@ARTICLE{Lattimer1989,
       author = {{Lattimer}, James M. and {Yahil}, A.},
        title = "{Analysis of the Neutrino Events from Supernova 1987A}",
      journal = {\apj},
         year = 1989,
        month = may,
       volume = {340},
        pages = {426},
          doi = {10.1086/167404},
       adsurl = {https://ui.adsabs.harvard.edu/abs/1989ApJ...340..426L}
}

@ARTICLE{Ginat2026,
       author = {{Ginat}, Yonadav Barry and {Antonini}, Fabio and {Flanagan}, Elizabeth and {Gieles}, Mark},
        title = "{Second-Generation Mass Peak in the Gravitational-Wave Population as a Probe of Globular Clusters}",
      journal = {arXiv e-prints},
         year = 2026,
        month = apr,
          eid = {arXiv:2604.07456},
        pages = {arXiv:2604.07456},
          doi = {10.48550/arXiv.2604.07456},
archivePrefix = {arXiv},
       eprint = {2604.07456},
 primaryClass = {astro-ph.HE},
       adsurl = {https://ui.adsabs.harvard.edu/abs/2026arXiv260407456G}
}

@article{Renaud2017,
  title = {The Origin of the {{Milky Way}} Globular Clusters},
  author = {Renaud, Florent and Agertz, Oscar and Gieles, Mark},
  year = 2017,
  month = mar,
  journal = {Monthly Notices of the Royal Astronomical Society},
  volume = {465},
  pages = {3622--3636},
  publisher = {OUP},
  doi = {10.1093/mnras/stw2969}
}

@article{Meng2022,
  title = {Tidal Disruption of Star Clusters in Galaxy Formation Simulations},
  author = {Meng, Xi and Gnedin, Oleg Y.},
  year = 2022,
  month = sep,
  journal = {Monthly Notices of the Royal Astronomical Society},
  volume = {515},
  pages = {1065--1077},
  publisher = {OUP},
  doi = {10.1093/mnras/stac1751}
}

@INPROCEEDINGS{Offner2023,
       author = {{Offner}, S.~S.~R. and {Moe}, M. and {Kratter}, K.~M. and {Sadavoy}, S.~I. and {Jensen}, E.~L.~N. and {Tobin}, J.~J.},
        title = "{The Origin and Evolution of Multiple Star Systems}",
    booktitle = {Protostars and Planets VII},
         year = 2023,
       editor = {{Inutsuka}, S. and {Aikawa}, Y. and {Muto}, T. and {Tomida}, K. and {Tamura}, M.},
       series = {Astronomical Society of the Pacific Conference Series},
       volume = {534},
        month = jul,
        pages = {275},
          doi = {10.48550/arXiv.2203.10066},
archivePrefix = {arXiv},
       eprint = {2203.10066},
 primaryClass = {astro-ph.SR},
       adsurl = {https://ui.adsabs.harvard.edu/abs/2023ASPC..534..275O}
}

@ARTICLE{Sana2013,
       author = {{Sana}, H. and {de Koter}, A. and {de Mink}, S.~E. and {Dunstall}, P.~R. and {Evans}, C.~J. and {H{\'e}nault-Brunet}, V. and {Ma{\'\i}z Apell{\'a}niz}, J. and {Ram{\'\i}rez-Agudelo}, O.~H. and {Taylor}, W.~D. and {Walborn}, N.~R. and {Clark}, J.~S. and {Crowther}, P.~A. and {Herrero}, A. and {Gieles}, M. and {Langer}, N. and {Lennon}, D.~J. and {Vink}, J.~S.},
        title = "{The VLT-FLAMES Tarantula Survey. VIII. Multiplicity properties of the O-type star population}",
      journal = {\aap},
         year = 2013,
        month = feb,
       volume = {550},
          eid = {A107},
        pages = {A107},
          doi = {10.1051/0004-6361/201219621},
archivePrefix = {arXiv},
       eprint = {1209.4638},
 primaryClass = {astro-ph.SR},
       adsurl = {https://ui.adsabs.harvard.edu/abs/2013A&A...550A.107S}
}

@ARTICLE{Sana2025,
       author = {{Sana}, H. and {Shenar}, T. and {Bodensteiner}, J. and {Britavskiy}, N. and {Langer}, N. and {Lennon}, D.~J. and {Mahy}, L. and {Mandel}, I. and {de Mink}, S.~E. and {Patrick}, L.~R. and {Villase{\~n}or}, J.~I. and {Dirickx}, M. and {Abdul-Masih}, M. and {Almeida}, L.~A. and {Backs}, F. and {Berlanas}, S.~R. and {Bernini-Peron}, M. and {Bowman}, D.~M. and {Bronner}, V.~A. and {Crowther}, P.~A. and {Deshmukh}, K. and {Evans}, C.~J. and {Fabry}, M. and {Gieles}, M. and {Gilkis}, A. and {Gonz{\'a}lez-Tor{\`a}}, G. and {Gr{\"a}fener}, G. and {G{\"o}tberg}, Y. and {Hawcroft}, C. and {H{\'e}nault-Brunet}, V. and {Herrero}, A. and {Holgado}, G. and {Izzard}, R.~G. and {de Koter}, A. and {Janssens}, S. and {Johnston}, C. and {Josiek}, J. and {Justham}, S. and {Kalari}, V.~M. and {Klencki}, J. and {Kub{\'a}t}, J. and {Kub{\'a}tov{\'a}}, B. and {Lefever}, R.~R. and {van Loon}, J. Th. and {Ludwig}, B. and {Mackey}, J. and {Ma{\'\i}z Apell{\'a}niz}, J. and {Maravelias}, G. and {Marchant}, P. and {Mazeh}, T. and {Menon}, A. and {Moe}, M. and {Najarro}, F. and {Oskinova}, L.~M. and {Ovadia}, R. and {Pauli}, D. and {Pawlak}, M. and {Ramachandran}, V. and {Renzo}, M. and {Rocha}, D.~F. and {Sander}, A.~A.~C. and {Schneider}, F.~R.~N. and {Schootemeijer}, A. and {Sch{\"o}sser}, E.~C. and {Sch{\"u}rmann}, C. and {Sen}, K. and {Shahaf}, S. and {Sim{\'o}n-D{\'\i}az}, S. and {van Son}, L.~A.~C. and {Stoop}, M. and {Toonen}, S. and {Tramper}, F. and {Valli}, R. and {Vigna-G{\'o}mez}, A. and {Vink}, J.~S. and {Wang}, C. and {Willcox}, R.},
        title = "{A high fraction of close massive binary stars at low metallicity}",
      journal = {Nature Astronomy},
         year = 2025,
        month = sep,
       volume = {9},
        pages = {1337-1346},
          doi = {10.1038/s41550-025-02610-x},
archivePrefix = {arXiv},
       eprint = {2509.12488},
 primaryClass = {astro-ph.SR},
       adsurl = {https://ui.adsabs.harvard.edu/abs/2025NatAs...9.1337S}
}

@ARTICLE{Varma2019,
       author = {{Varma}, Vijay and {Field}, Scott E. and {Scheel}, Mark A. and {Blackman}, Jonathan and {Gerosa}, Davide and {Stein}, Leo C. and {Kidder}, Lawrence E. and {Pfeiffer}, Harald P.},
        title = "{Surrogate models for precessing binary black hole simulations with unequal masses}",
      journal = {Physical Review Research},
         year = 2019,
        month = oct,
       volume = {1},
       number = {3},
          eid = {033015},
        pages = {033015},
          doi = {10.1103/PhysRevResearch.1.033015},
archivePrefix = {arXiv},
       eprint = {1905.09300},
 primaryClass = {gr-qc},
       adsurl = {https://ui.adsabs.harvard.edu/abs/2019PhRvR...1c3015V}
}

@ARTICLE{Guerrero2026,
       author = {{Guerrero}, Alexandra G. and {Zevin}, Michael and {Maclean}, Duncan B. and {Breivik}, Katelyn and {Rodriguez}, Carl L. and {Briel}, Max M. and {Holz}, Daniel E.},
        title = "{When the stars don't align: Investigating inconsistencies in binary black hole formation across population synthesis codes}",
      journal = {arXiv e-prints},
         year = 2026,
        month = aug,
          eid = {arXiv:2608.21609},
        pages = {arXiv:2608.21609},
          doi = {10.48550/arXiv.2608.21609},
archivePrefix = {arXiv},
       eprint = {2608.21609},
 primaryClass = {astro-ph.HE},
       adsurl = {https://ui.adsabs.harvard.edu/abs/2026arXiv260821609G}
}


    \appendix

    \setcounter{table}{0}
    \renewcommand{\thetable}{A\arabic{table}}


\section{Details of BH IFMR Model Parameters}


    \begin{table}[h]
    \centering
    \scriptsize
    \begin{tabular}{cccc}
        \toprule
        Model &Parameter & Value & Description \\
        \midrule
        \multirow{6}{*}{\uSSE} & \texttt{mdflag}    & 3   & Stellar wind mass loss prescription of \citet{Belczynski2010} \\
        & \multirow{2}{*}{\texttt{nsflag}}    & 3 (\uSSErapid) & \multirow{2}{*}{Remnant mass and fallback prescriptions of \citet{Fryer2012}} \\
        &                & 4 (\uSSEdelay)  & \\
        & \texttt{psflag}  & 1    & PPISN/PISN prescriptions of \citet{Belczynski2016} \\
        & \texttt{pts1},\texttt{2},\texttt{3} &  0.001, 0.02, 0.02 & Time step parameters \\
        & \texttt{mxns}    & 2.5 & Maximum mass of NS \\
        \midrule\midrule
        \multirow{6}{*}{\SEVN} & \texttt{tables}  &  \texttt{SEVNtracks\_parsec\_ov05\_AGB} &  Stellar evolution track lookup tables \\
        & \multirow{2}{*}{\texttt{snmode}}    & \texttt{rapid} (\SEVNrapid) & \multirow{2}{*}{Remnant mass and fallback prescriptions of \citet{Fryer2012}} \\
        &                & \texttt{delayed} (\SEVNdelay) & \\
        & \texttt{sn\_max\_ns\_mass} &  3   &  Maximum mass of NS \\
        & \texttt{sn\_pairinstability} & \texttt{mapelli20} & PPISN/PISN prescriptions of \citet{Mapelli2020} \\
        & \texttt{sn\_neutrinomloss}   & \texttt{lattimer89}  &   SN neutrino mass loss prescription of \citet{Lattimer1989} \\
        & \texttt{sn\_co\_lower\_sn}  &  1.44  &  Minimum \chem{CO} core mass to explode as SN \\
        \bottomrule
    \end{tabular}
    \caption{
        The list of flags and values used to compute the BH IFMR and fallback
        grids of both the \uSSE and \SEVN models, using the updated version of
        \SSE presented by \citet{Banerjee2020}%
        \footnote{\url{https://github.com/sambaranb/updated-BSE}} and
        version 2.15 of \SEVN\footnote{\url{https://sevncodes.gitlab.io/sevn/}}.
        Only flags which have an impact on BH formation
        are shown, as the WD and NS masses and the BH natal kicks are
        computed separately (see \Cref{sec:coupled_models,sec:bh_methods}), and
        no binary effects are included.
    }
    \label{table:sev_model_parameters}
    \end{table}


\end{document}